\documentclass[twoside,twocolumn,9pt]{article}
\usepackage{extsizes}
\usepackage[super,sort&compress,comma]{natbib} 
\usepackage[version=3]{mhchem}
\usepackage[left=1.5cm, right=1.5cm, top=1.785cm, bottom=2.0cm]{geometry}
\usepackage{balance}
\usepackage{mathptmx}
\usepackage{sectsty}
\usepackage{graphicx} 
\usepackage{lastpage}
\usepackage[format=plain,justification=justified,singlelinecheck=false,font={stretch=1.125,small,sf},labelfont=bf,labelsep=space]{caption}
\usepackage{float}
\usepackage{fancyhdr}
\usepackage{fnpos}
\usepackage[english]{babel}
\addto{\captionsenglish}{%
  
}
\usepackage{array}
\usepackage{droidsans}
\usepackage{charter}
\usepackage[T1]{fontenc}
\usepackage[usenames,dvipsnames]{xcolor}
\usepackage{setspace}
\usepackage[compact]{titlesec}
\usepackage{hyperref}

\usepackage[markup=underlined]{changes}

\definechangesauthor[color=Blue]{KS}
\definechangesauthor[color=Green]{JP}
\definechangesauthor[color=Orange]{AK}

\definecolor{addedblue}{RGB}{0, 0, 255} 
\definecolor{deletedred}{RGB}{255, 0, 0} 

\setaddedmarkup{\textcolor{addedblue}{#1}}
\setdeletedmarkup{\textcolor{deletedred}{\sout{#1}}}

\DeclareMathAlphabet{\mathcal}{OMS}{cmsy}{m}{n}

\definecolor{cream}{RGB}{222,217,201}

\begin{document}

\pagestyle{fancy}
\thispagestyle{plain}
\fancypagestyle{plain}{
\renewcommand{\headrulewidth}{0pt}
}

\makeFNbottom
\makeatletter
\renewcommand\LARGE{\@setfontsize\LARGE{15pt}{17}}
\renewcommand\Large{\@setfontsize\Large{12pt}{14}}
\renewcommand\large{\@setfontsize\large{10pt}{12}}
\renewcommand\footnotesize{\@setfontsize\footnotesize{7pt}{10}}
\makeatother

\renewcommand{\thefootnote}{\fnsymbol{footnote}}
\renewcommand\footnoterule{\vspace*{1pt}%
\color{cream}\hrule width 3.5in height 0.4pt \color{black}\vspace*{5pt}} 
\setcounter{secnumdepth}{5}

\makeatletter 
\renewcommand\@biblabel[1]{#1}            
\renewcommand\@makefntext[1]%
{\noindent\makebox[0pt][r]{\@thefnmark\,}#1}
\makeatother 
\renewcommand{\figurename}{\small{Fig.}~}
\sectionfont{\sffamily\Large}
\subsectionfont{\normalsize}
\subsubsectionfont{\bf}
\setstretch{1.125} 
\setlength{\skip\footins}{0.8cm}
\setlength{\footnotesep}{0.25cm}
\setlength{\jot}{10pt}
\titlespacing*{\section}{0pt}{4pt}{4pt}
\titlespacing*{\subsection}{0pt}{15pt}{1pt}

\fancyfoot{}
\fancyfoot[RO]{\footnotesize{\sffamily{1--\pageref{LastPage} ~\textbar  \hspace{2pt}\thepage}}}
\fancyfoot[LE]{\footnotesize{\sffamily{\thepage~\textbar\hspace{3.45cm} 1--\pageref{LastPage}}}}
\fancyhead{}
\renewcommand{\headrulewidth}{0pt} 
\renewcommand{\footrulewidth}{0pt}
\setlength{\arrayrulewidth}{1pt}
\setlength{\columnsep}{6.5mm}
\setlength\bibsep{1pt}

\makeatletter 
\newlength{\figrulesep} 
\setlength{\figrulesep}{0.5\textfloatsep} 

\newcommand{\topfigrule}{\vspace*{-1pt}%
\noindent{\color{cream}\rule[-\figrulesep]{\columnwidth}{1.5pt}} }

\newcommand{\botfigrule}{\vspace*{-2pt}%
\noindent{\color{cream}\rule[\figrulesep]{\columnwidth}{1.5pt}} }

\newcommand{\dblfigrule}{\vspace*{-1pt}%
\noindent{\color{cream}\rule[-\figrulesep]{\textwidth}{1.5pt}} }

\makeatother

\twocolumn[
  \begin{@twocolumnfalse}
\par
\vspace{1em}
\sffamily
\begin{tabular}{m{4.5cm} p{13.5cm} }

 & \noindent\LARGE{\textbf{When do molecular polaritons behave like optical filters?$^\dag$}} \\
\vspace{0.3cm} & \vspace{0.3cm} \\
 & \noindent\large{Kai Schwenncike,$^\ddagger$\textit{$^{a}$} Arghadip Koner,$^\ddagger$\textit{$^{a}$} Juan B. P{\'e}rez-S{\'a}nchez,$^\ddagger$\textit{$^{a}$} Wei Xiong,\textit{$^{a}$} Noel C. Giebink,\textit{$^{b}$} Marissa L. Weichman,\textit{$^{c}$} and Joel Yuen Zhou$^{\ast}$\textit{$^{a}$}} \\
 & \noindent\normalsize{This review outlines several linear optical effects featured by molecular polaritons arising in the collective strong light-matter coupling regime. Under weak laser irradiation and when the \textit{single-molecule} light-matter coupling can be neglected (often in the limit when the number of molecules per photon mode is large), we show that the excited-state molecular dynamics under collective strong coupling can be exactly replicated without the cavity using a shaped (or ``filtered'') laser, whose field amplitude is enhanced by the cavity quality factor, shining on the bare molecules. As a consequence, the absorption within a cavity can be understood as the overlap between the polariton transmission and the bare molecular absorption, suggesting that polaritons act in part as optical filters. This framework demystifies and provides a straightforward explanation for a large class of experiments and} theoretical models in molecular polaritonics, highlighting that the same effects can be achieved without the cavity with shaped laser pulses. With a few modifications, this simple conceptual picture can also be adapted to understand the incoherent nonlinear response of polaritonic systems. This review establishes a clear distinction between polaritonic phenomena that can be fully explained through classical linear optics and those that require a quantum electrodynamics approach. It also highlights the need to differentiate between effects that necessitate polaritons (i.e., hybrid light-matter states) and those that can occur in the weak coupling regime. We further discuss that certain quantum optical effects like fluorescence can be partially described as optical filtering, whereas some others like cavity-induced Raman scattering go beyond this. Further exploration in these areas is needed to uncover novel polaritonic phenomena beyond optical filtering.

\end{tabular}

 \end{@twocolumnfalse} \vspace{0.6cm}

  ]

\renewcommand*\rmdefault{bch}\normalfont\upshape
\rmfamily
\section*{}
\vspace{-1cm}


\footnotetext{\textit{$^{a}$~Department of Chemistry and Biochemistry, University of California, San Diego, La Jolla, CA 92093, USA.}}
\footnotetext{\textit{$^{b}$~Department of Electrical Engineering and Computer Science, and Physics, University of Michigan, Ann Arbor, MI 48109, USA. }}
\footnotetext{\textit{$^{c}$~Department of Chemistry, Princeton University, Princeton, NJ 08544, USA.}}

\footnotetext{\dag~Supplementary Information available: Estimating the number of molecules per photon mode in the collective strong coupling regime. See DOI: 00.0000/00000000.}
\footnotetext{$^\ddagger$~Co-authors.}



\section{Introduction}\label{sec1}

The interaction of matter with light drives numerous chemical, physical, and biological processes,
typically falling within the ``weak'' coupling regime in nature. Here,
light and matter are considered distinct entities where the former perturbatively
influences the latter. Weak coupling is defined by the rate
of energy exchange between light and matter being slow compared to
their natural dissipation rates; thus, well-defined light absorption and emission events can be identified. By confining the radiation field within a small mode volume, light and matter can coherently exchange energy multiple times before the photon escapes, giving rise to the ``strong'' coupling regime.
In this regime, light and matter form hybrid eigenmodes known as polaritons.

The hybridization of light and matter excitations in crystalline solids outside of cavities has been long known since the seminal works of Tolpygo~\cite{tolpygo1950physical} and Huang~\cite{huang1951lattice} on phonon-polaritons and Agranovich~\cite{agranovich1957influence} and Hopfield~\cite{hopfield1958theory} on exciton-polaritons in the 1950s. It wasn't, however, until 1992 that Weisbuch~\cite{weisbuch1992observation} demonstrated that these strong coupling phenomena in inorganic semiconductors could be enhanced with optical microcavities (though other work has since shown that the dielectric contrast between a bulk semiconductor and its surroundings can give rise to effective polariton formation even without the use of mirrors~\cite{canales2021abundance}). A few years later, Lidzey~\cite{lidzey1998strong} reported the same feat using excitons in disordered organic films to form microcavity polaritons. The study of strong coupling for ensembles of molecular vibrations is much more recent~\cite{shalabney2015coherent,long2015coherent},
yet serves as a testament to the ubiquity of polaritonic phenomena across the electromagnetic spectrum.

Today, the burgeoning field of molecular polaritonics shows promise in a diverse range
applications, such as manipulating chemical reactivity in both
ground~\cite{thomas2016ground,thomas2019tilting,campos2019resonant,du2022catalysis,ahn2023modification}
and excited states~\cite{hutchison2012modifying,herrera2016cavity, galego2016suppressing,galego2017many,mandal2019investigating,hoffmann2020effect,zeng2023control,dutta2024thermal,lee2024controlling}, enhancing exciton transport~\cite{rozenman2018long,pandya2022tuning,engelhardt2022unusual,engelhardt2023polariton,suyabatmaz2023vibrational,xu2023ultrafast,sokolovskii2023multi},
facilitating long-range energy transfer~\cite{coles2014polariton,zhong2016non,zhong2017energy,georgiou2018control,du2018theory,delpo2021polariton,son2022energy},
enabling room temperature polariton condensation~\cite{kasprzak2006bose,kena2010room,mazza2013microscopic,daskalakis2014nonlinear,plumhof2014room,grant2016efficient,pannir2022driving},
altering organic photophysical dynamics~\cite{stranius2018selective,martinez2018polariton,takahashi2019singlet,berghuis2019enhanced,polak2020manipulating,dovzhenko2021polariton,wu2022optical},
and modifying phase transitions~\cite{wang2014phase}. Despite these
advancements, the field is still marked by contradictory findings. For instance, multiple studies report no modification to chemical
reactivity due to polaritonic effects ~\cite{imperatore2021reproducibility,wiesehan2021negligible,thomas2023nonpolaritonic,fidler2023ultrafast}
and little to no polaritonic modification to spin conversion rates~\cite{eizner2019inverting,liu2020role}. Furthermore, with respect
to modification of phase transitions, recent reports find that the re-scaling of the local temperature inside the cavity with respect to the
temperature measured outside the cavity is the dominant mechanism, rather than the proposed renormalization of the material's free energy~\cite{jarc2023cavity,brawley2023sub}. This mix of positive and negative results underscores the nascent nature of the field and the need for
further experimental and theoretical understanding, as highlighted
by recent perspectives~\cite{wang2021roadmap,campos2023swinging,khazanov2023embrace}.

In an attempt to deepen this understanding at the recent 2023 Strong Coupling
with Organic Molecules (SCOM) Conference held in La Jolla, California, an important discussion
came into the limelight: the field must draw a clear divide between phenomena that can be understood solely through the language of classical optics, and those that go beyond this description.
Amidst this and similar discussions, a curious
revelation has emerged: consistently, classical linear optical descriptions like transfer matrix methods~\cite{zhu1990vacuum,schubert1996polarization,yariv2007photonics,nemet2020transfer}
elucidate many experimental findings~\cite{herrera2020molecular,simpkins2023control,wright2023rovibrational},
even in the realm of nonlinear spectroscopy~\cite{dunkelberger2016modified,wei2018nonlinear1,delpo2021polariton,renken2021untargeted,cheng2022molecular,hirschmann2023role}.
This raises the question: How can classical linear optics capture so many
phenomena attributed to molecular polaritons? Furthermore, does the success of the classical treatment imply that the language of cavity quantum-electrodynamics is unnecessary under certain circumstances relevant to polariton chemistry?

This review endeavors to provide a succinct yet comprehensive
account of a class of molecular polaritonic phenomena
that can be understood through the use of classical linear optics. First, a theoretical framework is established to understand the linear polaritonic spectra in the limit $N\to\infty$, or the large-$N$ limit, where $N$ is the number of molecules coupled to each photon mode (equivalently, when {single-molecule} light-matter coupling can be neglected). In this limit, the molecular-polariton absorption spectrum can be expressed as the overlap between the polariton transmission and bare molecular absorption spectra, resembling an optical-filtering behavior. In the regime where the polaritonic transmission and bare molecular absorption spectra are well separated in frequency, we make a connection to classical transfer matrix methods, demonstrating the agreement between the large-$N$ limit and classical linear optics. We then connect this frequency domain
picture to a dynamical perspective.

\begin{figure*}[!ht]
\begin{centering}
\includegraphics[width=\linewidth]{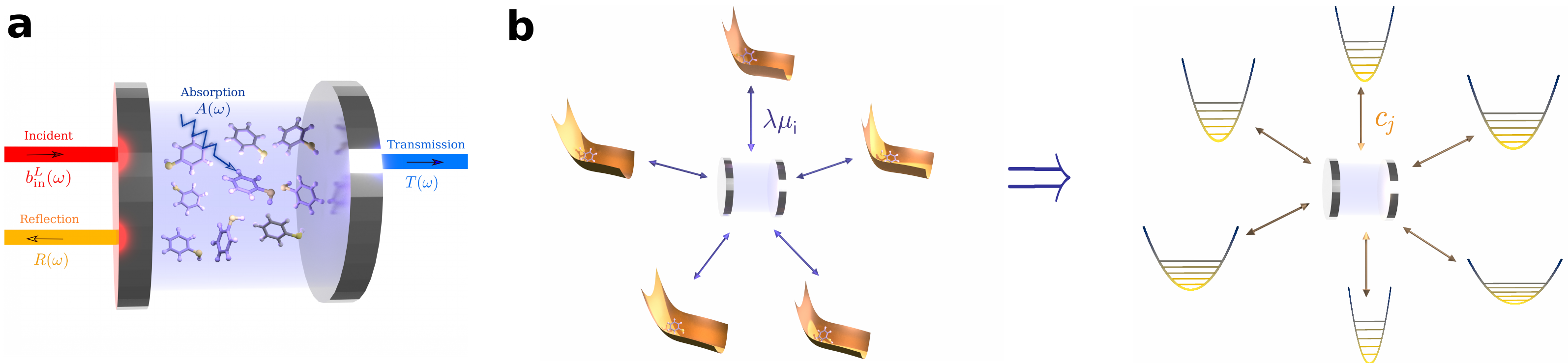}
\end{centering}
\captionof{figure}{\textit{Molecular polaritons and their linear spectroscopies.--} (a) Linear spectroscopy can report on the frequency-dependent transmission, absorption, and reflection of light by a cavity structure. (b) The molecular polariton problem -- where the anharmonic degrees of freedom of a large number of molecules, $N$, is coupled to a single (harmonic) photon mode -- can be regarded as a quantum impurity model. When $N\to\infty$, the reduced dynamics of the photon can be computed exactly by replacing the molecular degrees of freedom with a surrogate harmonic bath. (Reproduced from Ref.~\citenum{yuen2023linear}, Copyright 2024 J. Yuen-Zhou and A. Koner, published by  AIP Publishing.) \label{fig:process}}
\end{figure*}

In order to build upon the understanding of the optical filtering regime of molecular polaritonics, we provide explicit examples, both theoretical and experimental, where seemingly nontrivial polaritonic phenomena can be explained with concepts from linear optics. For instance, if we consider pumping polaritonic systems with a broadband laser pulse, we observe that both the relative steady-state populations and coherent molecular dynamics are identical to what we would observe pumping the bare molecular system outside of the cavity with the correctly shaped laser pulse whose intensity profile matches the polariton transmission spectrum.  We argue that under certain circumstances, the nonlinear optical response of many polaritonic systems can be understood via these optical filtering arguments, albeit with the caveat that the cavity transmission spectrum through which filtering occurs is itself a time-dependent function of the excitation dynamics of the intracavity molecules. These observations caution making claims of ``polaritonic'' effects before optical filtering effects have been judiciously accounted for.  We then discuss the limitations
of this approach, particularly in the limit of finite $N$ (\textit{i.e.}, few-molecule strong coupling) and in other situations that extend beyond a classical optics treatment of molecular polaritons. Furthermore, we provide a few existing experimental results in the collective strong coupling regime that cannot be accounted for by optical filtering, indicating ``non-trivial'' polaritonic effects.
\section{Theory}\label{theory}

\subsection{Exact solutions of the polariton Hamiltonian}\label{sec:impurity_model}
This section explores the scenario where $N$ molecules are coupled to a single harmonic cavity mode of frequency $\omega_{ph}$. The relevant Hamiltonian is the Tavis-Cummings Hamiltonian, extended to include the molecular vibrational degrees of freedom~\cite{tavis1968exact,perez2023simulating,yuen2023linear,feist2018polaritonic}:
\begin{align}
    H&=H_{0}+V, \label{eq:Hamiltonian}
\end{align}    
    where $H_{0}=H_{ph}+H_{mol}$ with,
\begin{align}\label{bare_cav_mol}
    H_{ph} =\hbar\omega_{ph}a^{\dagger}a,\quad\quad H_{mol}	=\sum_{i}^{N}H_{i}(q_{i},Q_{i}),
\end{align}      
as the zeroth-order bare cavity and molecular Hamiltonians, respectively, and 
\begin{align} \label{eq:dipolar_coupling}
    V=-\hbar\lambda(a+a^{\dagger})\mu
\end{align}
is the dipolar light-matter interaction. Here $a$ and $a^{\dagger}$ are the photon annihilation and creation operators that satisfy $[a,a^{\dagger}]=1$, and $H_{i}(q_{i}, Q_{i})$ is the molecular Hamiltonian for the $i$-th molecule, depending on the respective electronic $q_{i}$ and nuclear $Q_{i}$ degrees of freedom. In Eq. \ref{eq:dipolar_coupling}, $\mu=\sum_{i}\mu_{i}(q_{i},Q_{i})$ is the net molecular dipole operator and $\hbar\lambda=\sqrt{\frac{\hbar\omega_{ph}}{2\varepsilon_{0}\mathcal{V}_{ph}}}$ is the vacuum electric field. The scalar $\varepsilon_{0}$ is the vacuum permittivity, and $\mathcal{V}_{ph}$ is the cavity mode volume. Hereafter, for simplicity, we consider the case where the molecular sample occupies the same volume as the cavity mode, such that $\mathcal{V}_{mol}=\mathcal{V}_{ph}$.

Typically, the number of molecules per cavity mode is $N\approx10^{7}$ for electronic strong coupling~\cite{daskalakis2017polariton} and $N\approx10^{11}-10^{12}$ for vibrational strong coupling~\cite{del2015quantum} (see Section S1 in the Supplementary Information$^{\dag}$ for example calculations). Considering the photon mode as the \textit{system} and the molecular degrees of freedom as the \textit{bath}, the polariton setup can be mapped to a quantum impurity model with the \textit{impurity} (system) being coupled to a ``large'' environment. Makri~\cite{makri1999linear} has demonstrated that, in the limit of an infinite \textit{bath} size ($N\rightarrow \infty$), and initializing in an uncorrelated thermal system-bath state, $\rho=\rho_{ph}\otimes\rho_{mol}$, the \textit{reduced system dynamics} can be exactly obtained by replacing the complex anharmonic molecular bath with a surrogate bath of harmonic oscillators (see Fig.~\ref{fig:process}), $H_{mol}\rightarrow H_{mol}^{\text{eff}}=\sum_{j}\hbar\omega_{j}b_{j}^{\dagger}b_{j}$, coupled to the impurity (cavity mode), $V\rightarrow V^{\text{eff}}=-(a+a^{\dagger})\sum_{j}\hbar\bar{c}_{j}(b_{j}+b_{j}^{\dagger})$, where $\bar{c}_j$ are the respective effective couplings. 

This result was recently generalized to any arbitrary uncorrelated stationary (with respect to $H_{0}$) initial state (not necessarily thermal)~\cite{yuen2023linear}. This mapping is a consequence of the central limit theorem and relies on the fact that for $n > 2$, the $n$-th order bath correlation functions decay with increasing system size at a rate of $\mathcal{O}(N^{-1/2})$, given that the single molecule coupling strengths $\bar{c}_i = \mathcal{O}(N^{-1/2})$, a condition typically satisfied in the polariton problem~\cite{yuen2023linear}. These vanishing higher-order correlations essentially mean that the cavity probes only the linear response of the molecular bath. Owing to Fano, the linear response of an anharmonic $n$-level system can be decomposed into the response of independent harmonic oscillators at each transition frequency (see Eq.\ ~\ref{eq:linear_suceptability} below)~\cite{ fano1973impact,MukamelBook}. These transitions correspond to the harmonic frequencies contributing to the surrogate bath~\cite{yuen2023linear,gunasekaran2025continuum,cwik2016excitonic}.

\begin{table*}[!ht]
\caption{Molecular polariton linear spectra for arbitrary $N$ and $N\to\infty$\label{tab:Molecular-Polariton-Linear-Spectrum}}
\centering
\begin{tabular}{ccc}
Signal & Arbitrary $N$\textsuperscript{\textit{a}} & $N\to\infty$\tabularnewline
\hline 
\noalign{\vskip0.25cm}
$A(\omega)$ & $-\kappa_{L}\Big\{\kappa|D^{(R)}(\omega)|^{2}+2\textrm{Im} [D^{(R)}(\omega)]\Big\}$ & $\frac{\kappa_{L}\omega_{ph}\textrm{Im}[\chi(\omega)]}{|\omega-\omega_{ph}+i\frac{\kappa}{2}+\frac{\omega_{ph}}{2}\chi(\omega)|^{2}}$\tabularnewline
\noalign{\vskip0.25cm}
$T(\omega)$ & $\kappa_{L}\kappa_{R}|D^{(R)}(\omega)|^{2}$ & $\frac{\kappa_{L}\kappa_{R}}{|\omega-\omega_{ph}+i\frac{\kappa}{2}+\frac{\omega_{ph}}{2}\chi(\omega)|^{2}}$\tabularnewline
\noalign{\vskip0.25cm}
$R(\omega)$ & $1+2\kappa_{L}\textrm{Im} [D^{(R)}(\omega)]+\kappa_{L}^{2}|D^{(R)}(\omega)|^{2}$ & $1-\frac{\kappa_{L}\Big\{\kappa_{R}+\omega_{ph}\textrm{Im}[\chi(\omega)]\Big\}}{|\omega-\omega_{ph}+i\frac{\kappa}{2}+\frac{\omega_{ph}}{2}\chi(\omega)|^{2}}$\tabularnewline
\noalign{\vskip0.25cm}
\hline 

\multicolumn{3}{c}{\footnotesize \textsuperscript{\textit{a}}Here $D^{(R)}(\omega)=-i\int_{-\infty}^{\infty}dte^{i\omega t}\Theta(t)\langle[a(t),a^{\dagger}]\rangle$
is the cavity retarded Green's function.}\tabularnewline
\end{tabular}
\end{table*}

\subsection{Spectroscopic observables from the photon Green's function}

The standard polaritonic linear spectroscopic observables can be expressed in terms of the retarded Green's function of the cavity, $D^{(R)}(\omega)=-i\int_{-\infty}^{\infty}dte^{i\omega t}\Theta(t)\langle[a(t),a^{\dagger}]\rangle$ (see Table \ref{tab:Molecular-Polariton-Linear-Spectrum}), where $\langle \dots \rangle$ denotes a trace with respect to the initial system bath density matrix. In the limit $N\to \infty$, the mapping discussed above reduces the polariton problem into one of linearly coupled harmonic oscillators: the originally harmonic cavity photon mode couples to the surrogate harmonic bath. This leads to readily obtainable analytical solutions~\cite{yuen2023linear,cwik2016excitonic,zeb2018exact, schwennicke2024extracting}.

In the limit $N\rightarrow\infty$, the frequency-resolved absorption, transmission, and reflection in the first excitation manifold, under the rotating wave approximation (RWA) in the cavity-laser interaction, are given by~\cite{yuen2023linear}: 
\begin{subequations}\label{spectra_formula}
\begin{align}
    A(\omega) & =\frac{\kappa_{L}\omega_{ph}\textrm{Im}[\chi^{(1)}(\omega)]}{|\omega-\omega_{ph}+i\frac{\kappa}{2}+\frac{\omega_{ph}}{2}\chi^{(1)}(\omega)|^{2}},\label{eq:absorption}\\
    T(\omega) & =\frac{\kappa_{L}\kappa_{R}}{|\omega-\omega_{ph}+i\frac{\kappa}{2}+\frac{\omega_{ph}}{2}\chi^{(1)}(\omega)|^{2}},\label{eq:transmission}\\
    R(\omega) & =1-A(\omega)-T(\omega).\label{eq:reflection}
\end{align}
\end{subequations}
Here the cavity linewidth is $\kappa=\kappa_{L}+\kappa_{R}$, where $\kappa_{L}$ and $\kappa_{R}$ denote the cavity decay rates into the left and right photon continua. The molecular linear
susceptibility~\cite{MukamelBook} is given by
\begin{equation}
\chi^{(1)}(\omega)=\lim_{\gamma\to 0^{+}}\frac{1}{\hbar\varepsilon_{0}\mathcal{V}_{mol}}\sum_{y,z}[p_{y}-p_{z}]\frac{|\langle z|\mu|y\rangle|^{2}}{\omega_{zy}-\omega-i\frac{\gamma}{2}}\label{eq:linear_suceptability}
\end{equation}
where $\mathcal{V}_{mol}$ is the volume of the molecular sample, $|y\rangle$ and $|z\rangle$ are the many-body eigenstates
of $H_{mol}$ in Eq.\ \ref{bare_cav_mol}, and $p_{y}$ and $p_{z}$ are
the corresponding initial populations. Note that the damping coefficient $\gamma \to 0^+$ is introduced to ensure causality. It is important to comment here that although Eqs.~\ref{eq:absorption}, \ref{eq:transmission}, and \ref{eq:reflection} assume the RWA in the cavity-laser interaction and a single photon mode interacting with the molecules, the implications hold beyond the RWA and for a multimode cavity as \'Cwik \textit{et al.}~\cite{cwik2016excitonic} imply. The validity of this formalism for a multimode cavity relies on the fact that, in the regime that $N\rightarrow\infty$, the different cavity modes do not couple to one another~\cite{cwik2016excitonic,engelhardt2023polariton}, allowing for mode-wise separation of the photon Green's function, $D^{(R)}_k(\omega)$. In situations where the large-$N$ limit cannot be used to describe the polaritonic system, mode-mixing forbids a single mode description of the cavity.~\cite{Lindoy2024investigating}

The formalism presented above is significant given that, over the years, a variety of computational methods have been developed for the simulation of the polariton spectra. These range from quantum optical methods based on input-output theory that require simplification of the rich molecular complexity~\cite{reitz2019langevin,Li2018,Ribeiro2018vp,kansanen2021polariton} to models incorporating \textit{ab initio} treatments but limited to only a single or very few molecules~\cite{leppala2023linear}. Eqs.~\ref{eq:absorption}-\ref{eq:reflection} demonstrate that linear polaritonic spectra can be computed directly using the molecular linear susceptibility of the bare molecule outside the cavity, avoiding explicit simulations with $N$ molecules coupled to a photon mode. This idea is reminiscent of classical optical transfer matrix methods, which we present in the following section. 

\subsection{Connection to classical linear optics: Transfer matrix methods}\label{TMM}

The simplified dependence on the molecular linear susceptibility within
Eqs. \ref{eq:absorption}-\ref{eq:reflection} parallels classical
optics treatments, particularly the transfer matrix method (TMM)~\cite{zhu1990vacuum,schubert1996polarization,yariv2007photonics}. Using TMM, the following expression can be derived which is typically used to describe the fraction of light at frequency $\omega$ transmitted through a Fabry-Perot
cavity of length $\mathcal{L}$ containing a dielectric molecular sample:
\begin{equation}
T(\omega)=\frac{T_{L}T_{R}e^{-\alpha(\omega)\mathcal{L}}}{|1-\sqrt{R_{L}R_{R}}e^{-\alpha(\omega)\mathcal{L}}e^{i\frac{2\omega n(\omega)\mathcal{L}}{c}}|^2},\label{eq:Transmission_TMM}
\end{equation}
where $T_{L,R}$ and $R_{L,R}$ are the transmission and reflection coefficients
of the left and right mirrors, $\alpha(\omega)$ and $n(\omega)$
are the frequency-dependent absorption coefficient and refractive
index for the intracavity molecular sample, and $c$ is the speed of light. Assuming that the highly reflective cavity mirrors are
lossless (\textit{e.g.}, $R_{L,R}\gg T_{L,R}$ and $T_{L,R} + R_{L,R} = 1$) and the molecular absorption  decays rapidly away from the molecular resonance, the classical expression for transmission
near the polariton frequencies approximates to: 
\begin{small}
\begin{equation}
T(\omega)\approx\frac{T_{L}T_{R}\omega_{FSR}{}^{2}}{4\pi^{2}|\omega-m\omega_{FSR}+i\frac{\omega_{FSR}}{4\pi}(T_{L}+T_{R})+\frac{\omega\varepsilon_0}{2\varepsilon_B}\chi^{(1)}(\omega)|^{2}},\label{eq:Transmission_TMM_approx}
\end{equation}
\end{small}where $\varepsilon_B$ is the permittivity of the nonresonant background~\cite{khitrova1999nonlinear,gunasekaran2025continuum}, $\omega_{FSR}=\frac{\pi c}{\mathcal{L}n_B}$ is the cavity free spectral range (here $n_B=\textrm{Re}[\sqrt{\varepsilon_B/\varepsilon_0}]$ is the background refractive index), $m$ is a positive
integer representing the longitudinal cavity mode order, and $\chi^{(1)}(\omega)=[n(\omega)+i\frac{1}{2}\frac{c}{\omega}\alpha(\omega)]^{2}-\frac{\varepsilon_B}{\varepsilon_0}$
is the linear susceptibility of the molecular sample. By taking $\frac{T_{L,R}\omega_{FSR}}{2\pi}\to\kappa_{L,R}$,
 $m\omega_{FSR}\to\omega_{ph}$, $\frac{\omega}{2}\chi^{(1)}(\omega)\to\frac{\omega_{ph}}{2}\chi^{(1)}(\omega)$, and assuming no background (\textit{i.e.}, $\varepsilon_B=\varepsilon_0$),
the quantum mechanical expression emerges (compare to Eq.\ \ref{eq:transmission}).
As discussed in Section \ref{sec:impurity_model}, in the large-$N$ limit the anharmonic molecular bath can be treated using an \textit{effective} harmonic model, leading to the convergence between the quantum and classical treatment. This result is consistent with, and extends beyond, previous works that treated the molecular degrees of freedom as classical oscillators to calculate the polaritonic spectrum. \cite{rudin1999oscillator,torma2015strong, Lieberherr2023vibrational,climent2024kubo} Therefore, the agreement between the classical and the quantum treatment in the large-$N$ limit explains the success of TMM and other classical linear optics methods at modeling the molecular polariton spectra in the collective strong coupling regime.

\subsection{Polariton linear absorption as optical filtering}\label{Sec:optical_filtering_1}

\begin{figure}[!h]
\centering
\includegraphics[width=0.5\linewidth]{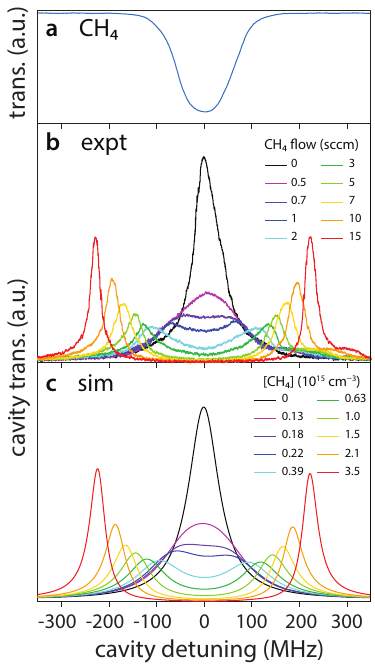}

\captionof{figure}{\textit{Linear spectroscopy of polaritons in gas-phase methane (CH$_4$).--} (a) Experimental transmission spectrum through a buffer gas cooled CH$_4$ sample showing absorption of light by the rovibrational transition at 3057.687423 cm$^{-1}$. The frequency axis is referenced with this transition corresponding to 0 MHz. (b) Experimental transmission spectrum of an $8.36$ cm long near-confocal Fabry–Pérot cavity containing CH$_4$ under systematic tuning of the intracavity gas flow rate. In all traces, a cavity mode is kept locked on resonance with the CH$_4$ transition of interest. (c) Simulated cavity transmission spectra obtained by fitting the corresponding experimental trace with the classical cavity transmission expression.  (Reproduced from Ref.~\citenum{wright2023rovibrational} with permission from the American Chemical Society, Copyright 2023.\label{fig:ch4})}
\end{figure}

\begin{figure*}[!ht]
\centering
\includegraphics[width=1\linewidth]{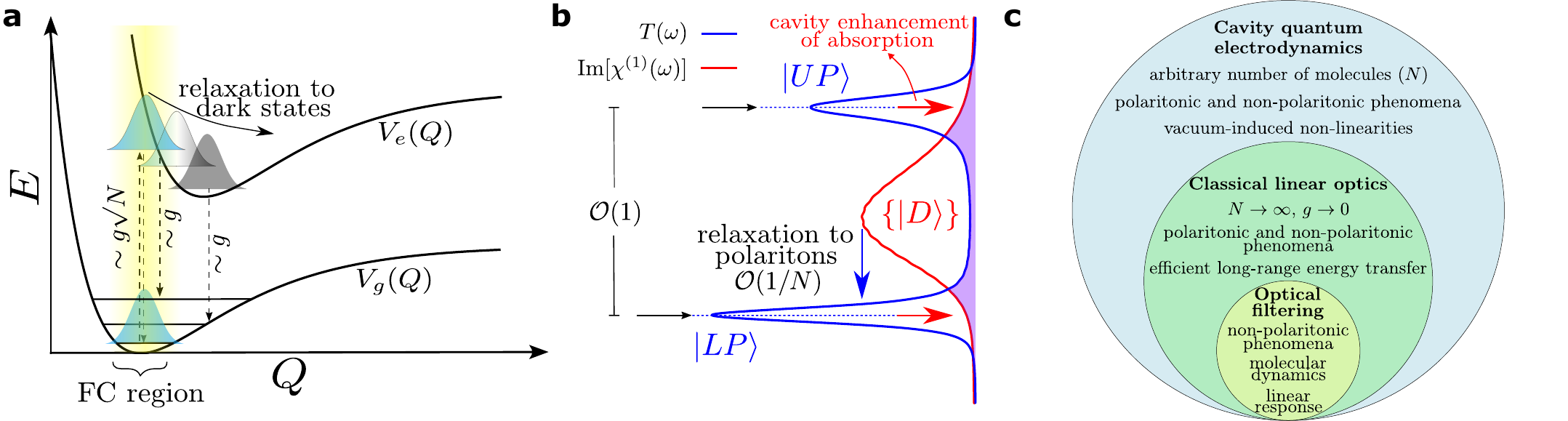}

\captionof{figure}{\textit{Collective vs.\ single-molecule coupling effects.}-- (a) Light-matter coupling is collectively enhanced for optical transitions that do not create vibrational excitations, \textit{i.e.}, phonons, in the ground state. This implies that only Rayleigh-like processes are relevant at short timescales $\sim{2\pi/g}$ for large $N$ ($\equiv$ small $g$). On the other hand, the light-matter coupling is single-molecule-like for processes that create phonons in the ground state, \textit{e.g.}, Raman scattering~\cite{koner2024nonlinear} and fluorescence~\cite{Perez2025radiative}. (b) Collective coupling creates upper and lower polaritons. Decay from polaritons to dark states occurs due to the loss of exciton coherence (vibrational dynamics away from the Franck-Condon (FC) region), and can be thought of as absorption through the polariton windows. Decay from dark states back to polariton states relies on the single-molecule light-matter coupling $g$, hence $\mathcal{O}(1/N)$. These processes are enhanced by the cavity-quality factor $\mathcal{Q}$. (c) Venn diagram illustrating the scenarios that require a quantum electrodynamics, a classical linear optics, and an optical filtering description. We also provide some phenomena discussed in this manuscript within each domain. The diagram is restricted to the first excitation manifold. (Fig. 3a adapted from Ref.~\citenum{perez2023simulating}, Copyright 2023 J. Pérez-Sánchez, et al., published by PNAS, and Ref.~\citenum{perez2023frequency}, Copyright 2024, J. Pérez-Sánchez, et al., published by the American Physical Society. Licensed under CC-BY 4.0. To view a copy of this license, visit https://creativecommons.org/licenses/by/4.0/.)\label{fig:cute}}
\end{figure*}
A closer inspection of Eqs. \ref{eq:absorption} and \ref{eq:transmission} reveals an intuitive relation,
\begin{equation}
A(\omega)=\frac{\omega_{ph}}{\kappa_{R}}\textrm{Im}[\chi^{(1)}(\omega)] T(\omega). \label{eq:absorption_reformulation}
\end{equation}
This equation shows that in the collective regime absorption within the cavity is directly proportional to the product of the bare, free-space molecular absorption $\textrm{Im}[\chi^{(1)}(\omega)]$ and the polaritonic cavity transmission spectrum $T(\omega)$. Note that the pre-factor $\frac{\omega_{ph}}{\kappa_{R}}$ -- which is a proportional to the cavity quality factor $\mathcal{Q}=\frac{\omega_{ph}}{\kappa}$ -- encodes cavity-enhancement of the molecular absorption. In a high-$\mathcal{Q}$ cavity constructed with highly-reflective mirrors, the photonic decay rate $\kappa_{R}$ will be small and the fraction of light absorbed by intracavity molecules will increase accordingly, as is well known in cavity-enhanced spectroscopy.~\cite{romanini2014introduction}

Viewed through the lens of linear optics, the polaritonic transmission peaks function in part as ``optical filters''~\cite{dutta2024thermal}, only allowing radiation of select frequencies outside to enter the cavity, where the molecules absorb based on their bare absorption spectrum. This picture is consistent with the work of Groenhof \textit{et al.},~\cite{groenhof2019tracking,dutta2024thermal} where the authors used molecular dynamics calculations to show that the population transfer between polaritons and dark states is determined by the overlap between the polaritonic and molecular absorption spectra.

We illustrate one experimental example of these effects in Fig.\ \ref{fig:ch4}, where  Rabi splittings in the transmission spectra of a Fabry–Pérot cavity containing various intracavity number densities of gas-phase methane (CH$_4$) are nearly perfectly captured using Eq.~\ref{eq:Transmission_TMM}.~\cite{wright2023rovibrational,wright2023versatile} At intermediate CH$_4$ number densities, the Rabi splitting does not significantly exceed the molecular linewidth (Fig.\ \ref{fig:ch4}a), and the polariton modes therefore have significant spectral overlap with the bare molecular absorption spectrum leading to their attenuation. The transmitted polariton intensity increases as the Rabi splitting increases, and the polariton bands move further out of range of the bare molecular absorption band.

\subsection{The quantum dynamics perspective}\label{sec:cute}

To understand where the linear optics picture breaks down, we must examine the $N\rightarrow\infty$ limit at the core of the derivation of Eq.~\ref{eq:absorption_reformulation}. Here we do so with the Collective dynamics Using Truncated Equations (CUT-E) formalism~\cite{perez2023simulating}, which is consistent with previous works by Spano \textit{et al.} in the context of J-aggregates~\cite{spano2011vibronic}, and works by Nitzan \textit{et al.}~\cite{Cui2022JCP} and Keeling \textit{et al.}~\cite{zeb2018exact,Piper2022PRL} in the context of molecular polaritons.

CUT-E is a decays representation of the Hamiltonian in a basis of coherent and incoherent excitons interacting with the cavity mode. The most important conclusion from CUT-E is that optical transitions that do not create phonons in the electronic ground state occur via the collective light-matter coupling component, while optical processes that create phonons in the electronic ground state are single-molecule like. Therefore, collective light-matter processes (e.g., Rabi oscillations) correspond to Rayleigh scattering, while vacuum-induced Raman scattering and fluorescence are of $\mathcal{O}(1/N)$ (see Fig.~\ref{fig:cute}a). This is reminiscent of how resonant Rayleigh scattering dominates the observed secondary radiation for a single quantum well strongly coupled to a semiconductor microcavity at early times, while photoluminescence dominates at longer times~\cite{hayes1998resonant}. This picture allows the separation of the light-matter coupling term into its collective ($g\sqrt{N}$) and single-molecule ($g=\lambda\mu/\sqrt{N}$) components~\cite{perez2023simulating,perez2023frequency,Perez2025cute}. In the large-$N$ limit, the latter can be treated perturbatively. The resulting zeroth-order Hamiltonian incorporates the collective light-matter coupling component and local vibrational dynamics, which is sufficient to describe the formation of polaritons and decay to dark states. On the other hand, the perturbation leads to relaxation from dark states back to polariton states (which is only relevant at longer timescales)\cite{Perez2025radiative} and cavity mediated Raman processes (which are negligible compared to absorption and other linear optical features) \cite{koner2024nonlinear}. Notably, this structure is preserved even in the presence of disorder (dCUT-E)~\cite{perez2023frequency}.
\begin{figure*}[!ht]
\centering
\includegraphics[scale=.9]{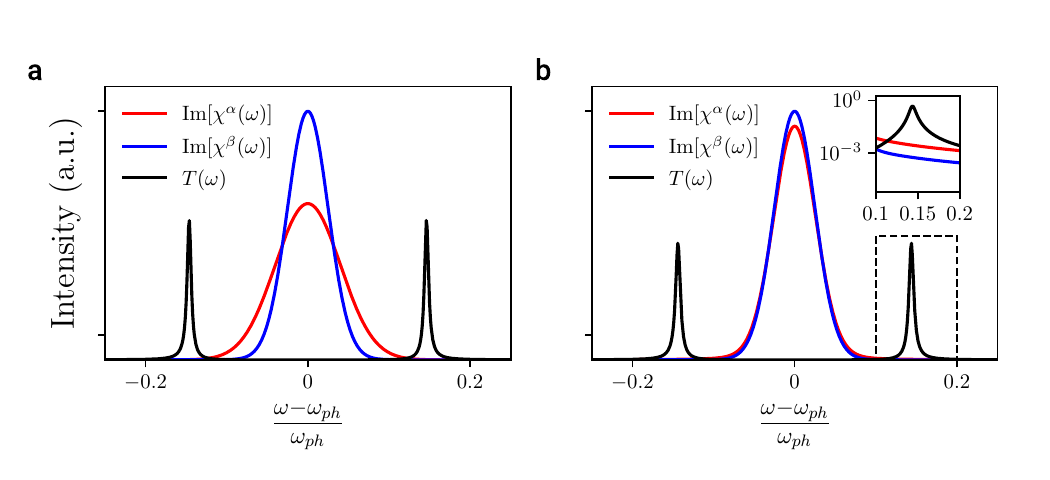}
\caption{\textit{Demystifying cavity-mediated ``non-statistical'' energy redistribution.--} Molecular species $\alpha$ and $\beta$ are coupled to a common cavity mode with the same coupling strength.  (a) Here, the homogeneous molecular linewidths are the same, while the inhomogeneous linewidth for $\alpha$ is significantly larger than that of $\beta$. Due to this inhomogeneity, $\alpha$ has an increased spectral overlap with the polariton transmission spectrum $T(\omega)$. As a result, the ratio of the energy absorbed by $\alpha$ over that absorbed by $\beta$ is greater than 1 ($\Delta \mathcal{E}^{\alpha}/\Delta \mathcal{E}^{\beta}=2.2$).
(b) Here the molecular species have the same inhomogeneous linewidth, but $\alpha$ possesses a larger homogeneous linewidth than $\beta$. Despite the superficial similarity of the absorption bands of the two molecular species outside of the cavity due to the predominant inhomogeneous broadening, we find
that $\Delta \mathcal{E}^{\alpha}/\Delta \mathcal{E}^{\beta}=2.7$. The log-scale inset highlights that while the centers of the molecular profiles are similar, $\alpha$ has considerably longer Lorentzian tails, still leading to improved spectral overlap with $T(\omega)$. Together, these examples illustrate that the tails of the molecular lineshapes matter significantly as their spectral overlap with the polariton windows determines intracavity energy absorption. \label{fig:Non-statistical-energy-redistribution}}
\end{figure*}

The collective coupling $g\sqrt{N}$ can be experimentally measured~\cite{schwennicke2024extracting}; therefore, in our analysis, we treat it as a finite constant. A consequence of this constraint is that taking the limit $N\rightarrow\infty$ is equivalent to taking $g\rightarrow 0$. In this limit, the previously mentioned $\mathcal{O}(1/N)$ non-linearities can be neglected. Hence only the molecular linear susceptibility $\chi^{(1)}(\omega)$ -- which is proportional to the two-point correlation function of the molecules and is therefore harmonic -- contributes to the polaritonic response (see Section~\ref{sec:impurity_model})~\cite{yuen2023linear}. This approximation, referred to as zeroth-order CUT-E, is valid for Eqs.~\ref{eq:absorption}-\ref{eq:reflection} since the photon-photon correlation function decays in an ultrafast timescale before fluorescence takes place ($\sim$ ns for the UV-visible and $\sim$ ms for the IR), and Raman and higher order scattering processes are weak compared to absorption. Other phenomena such as ultrafast polariton photochemistry should also be well approximated within this limit for the most part~\cite{dutta2024thermal,perez2023frequency}. At long times when $\mathcal{O}(1/N)$ processes begin to play an important role, the $N\rightarrow\infty$ approximation is no longer valid and quantum electrodynamics becomes relevant. Examples of these ``$1/N$ effects'' are relaxation from dark states back to polariton states via radiative pumping and vibrational relaxation (see Fig.~\ref{fig:cute}b). In Sec. \ref{sec:beyond_filtering} we discuss how these quantum effects are modified by collective strong coupling and when they can also be described as optical filtering.



\section{Examples of polaritonic effects interpretable as optical filtering}\label{examples}

In this section, we employ a linear optics viewpoint, in the $N\to\infty$ limit, to address specific instances of cavity-mediated non-statistical energy redistribution, control of photoreactivity via polaron decoupling, and coherent molecular excited state dynamics. These topics have been discussed in the literature as polaritonic phenomena of interest. While we do not question the correctness of prior literature reports, we seek here to demystify certain examples of these processes in light of the insights discussed above in Section 2. In each instance, we consider the $N\to\infty$ limit, enabling direct application of Eq.\ \ref{eq:absorption_reformulation}. As such, we predict that some polaritonic phenomena can be reproduced outside of a cavity by pumping with a laser field that has an intensity profile $|E(\omega)|^2$ proportional to the polariton transmission spectrum $T(\omega)$. As discussed in Section 2, Eq.\ \ref{eq:absorption_reformulation} pertains not only to steady-state molecular populations but also to excited state dynamics before the non-linear $\mathcal{O}(1/N)$ processes begin to matter. We insist, however, that these findings do not consider dynamics at times larger than $1/g$ nor when strong coupling is attained with small $N$. We will discuss these caveats, and others, in subsequent sections. Finally, at the end of this section, we review experimental examples of polaritonic effects that can be interpretable as optical filtering. 

\subsection{Non-statistical energy redistribution: A spectral overlap interpretation}\label{sec:ex1}

In this example, we consider two molecular species simultaneously coupled to a single cavity mode. Upon broadband optical excitation of the cavity, energy is observed to funnel selectively into one of the species with a ``non-statistical'' energy redistribution that differs from what one would expect if the same broadband excitation acted directly on the molecules in free space. This phenomenon was originally studied with computational rigor by Groenhof and Toppari~\cite{groenhof2018coherent} for a scenario involving up to 1,000 molecules of one species, and revisited by P{\'e}rez-S{\'a}nchez \textit{et al.}~\cite{perez2023simulating} for the case where the intracavity number of both molecular species tends to infinity. The intriguing observation in both studies is that the excitation energy is eventually funneled selectively into the molecular species that features the fastest dephasing. In Reference \citenum{perez2023simulating} P{\'e}rez-S{\'a}nchez \textit{et al.} explained this phenomenon with a time-domain interpretation: the cavity and the molecules exchange energy, and the molecules that dephase the fastest absorb energy at a faster rate and return less energy back to the cavity with every Rabi oscillation. There is no mistake in this time-domain interpretation, but here we argue that a much simpler frequency-domain picture can clarify the observed energy funneling. In short, no matter how complex the time-domain Rabi oscillations between molecules and cavity mode appear, they simply correspond to the frequency-domain filtering of broadband light through the polaritonic transmission windows. Temporal transients are immaterial if one is interested in steady-state observables such as the energy redistribution into different molecular species.  

For two distinct, non-interacting molecular species $\alpha$ and $\beta$ coupled to the same cavity
mode, Eq.\ \ref{eq:absorption_reformulation} gives the collective intracavity molecular absorption as:
\begin{equation}
A(\omega)=A^{(\alpha)}(\omega)+A^{(\beta)}(\omega)\label{eq:two_molecular_species_absorption},
\end{equation}
where $A^{(\alpha,\beta)}(\omega)=\frac{\omega_{ph}}{\kappa_{R}}\textrm{Im}[\chi^{(1\alpha,\beta)}(\omega)]T(\omega)$
is the fraction of light absorbed by molecular species $\alpha,\beta$ in the cavity. This equation clarifies the origin of the non-statistical energy redistribution, or funneling, for this system: the molecular species exhibiting the largest overlap between its bare absorption spectrum and the polariton transmission peaks absorbs the most energy.

Fig.\ \ref{fig:Non-statistical-energy-redistribution} showcases two
instructive scenarios that highlight the connection between
the bare molecular spectral overlap with the polariton transmission spectrum
and the ratio of the total energy absorbed by molecular species
$\alpha$ ($\Delta \mathcal{E}^{\alpha}$) to the total energy absorbed
by molecular species $\beta$ ($\Delta \mathcal{E}^{\beta}$) upon broadband excitation:
\begin{equation}
\frac{\Delta \mathcal{E}^{\alpha}}{\Delta \mathcal{E}^{\beta}}=\frac{\int d\omega \textrm{Im}[\chi^{(1\alpha)}(\omega)]T(\omega)}{\int d\omega \textrm{Im}[\chi^{(1\beta)}(\omega)]T(\omega)}.\label{eq:NA_NB}
\end{equation}
 In both instances, the bare molecular spectra are modeled as Voigt profiles centered at the cavity frequency $\omega_{ph}$:
 \begin{align}
 V(\omega;\sigma^{\alpha,\beta},\Gamma^{\alpha,\beta})=&\frac{1}{\pi}\int^\infty_{-\infty}d\omega^\prime \Big\{\frac{\exp\Big[-\Big(\frac{\omega^\prime}{\sqrt{2}\sigma^{\alpha,\beta}}\Big)^2\Big]}{\sigma^{\alpha,\beta}\sqrt{2\pi}} \nonumber \\
 &\times \frac{(\Gamma^{\alpha,\beta}/2)^2}{(\omega-\omega_{ph}-\omega^\prime)^2+(\Gamma^{\alpha,\beta}/2)^2}\Big\},
 \end{align} 
 where $\sigma^{\alpha,\beta}$ is the Gaussian (inhomogeneous) linewidth and $\Gamma^{\alpha,\beta}$ is the Lorenztian (homogeneous) linewidth for each molecular species. Species $\alpha$ and $\beta$ are coupled to a common cavity mode with the same coupling strength $g\sqrt{N}/\omega_{ph}=0.1$ and the cavity decay rate is taken to be $\kappa/\omega_{ph}=2\kappa_{R}/\omega_{ph}=2\kappa_{L}/\omega_{ph}=0.01$.  Fig.\ \ref{fig:Non-statistical-energy-redistribution}a presents the scenario where both species have the same homogeneous broadening ($\Gamma^{\alpha,\beta}/\omega_{ph}=0.001$), but species $\alpha$ has more significant inhomogeneous broadening ($\sigma^{\alpha}/\omega_{ph}=0.04$) compared to $\beta$ ($\sigma^{\beta}/\omega_{ph}=0.025$). Meanwhile, Fig.\ \ref{fig:Non-statistical-energy-redistribution}b presents the scenario where $\alpha$ and $\beta$ have the same inhomogeneous broadening ($\sigma^{\alpha,\beta}=0.025$), but $\alpha$ has larger homogeneous  broadening ($\Gamma^{\alpha}/\omega_{ph}=0.005$) than $\beta$ ($\Gamma^{\beta}/\omega_{ph}=0.001$). In both scenarios, the bare molecular spectrum of $\alpha$ possesses a larger spectral overlap with the polariton transmission spectrum than that of $\beta$; thus, molecular species $\alpha$ exhibits greater energy absorption. Notably, Fig.\ \ref{fig:Non-statistical-energy-redistribution}b emphasizes the significance of the Lorentzian tails of the linear molecular absorption spectrum in this energy absorption process (see inset). Therefore, while
scenarios (a) and (b) may seem distinct, their outcomes are explained by the same underlying mechanism. Again, the polariton peaks function as optical filters, determining the specific frequencies of light that are permitted into the cavity. Consequently, the species whose free space absorption spectrum overlaps more substantially with these frequencies absorbs the greater amount of energy. In fact, if a light source mimicking the intensity profile of the polariton transmission spectrum is employed outside of the cavity, the ratio of the energy absorbed by the two molecular species should be the same in free space as it would be under broadband illumination of the polaritonic system:
\begin{equation}
\frac{\Delta \mathcal{E}^{\alpha}}{\Delta \mathcal{E}^{\beta}}=\frac{\int d\omega \, \omega\textrm{Im}[\chi^{(1\alpha)}(\omega)]|E(\omega)|^{2}}{\int d\omega \, \omega\textrm{Im}[\chi^{(1\beta)}(\omega)]|E(\omega)|^{2}},\label{eq:outside_cav_filter}
\end{equation}
 where $|E(\omega)|^{2}\propto\frac{T(\omega)}{\omega}$ is the intensity profile of the shaped electromagnetic field. Note that this is true so long as the intensity of the driving field lies within the linear regime such that nonlinear effects are negligible. 
 
 This linear optics perspective offers a unified framework for interpreting previous theoretical findings. In the work of Groenhof and Toppari,~\cite{groenhof2018coherent} we can explain the preferential energy channeling to hydroxyphenyl-benzothiazole (HBT) over rhodamine (Rho) when the lower polariton is excited: the bare HBT linear absorption has more substantial overlap with the lower polariton than that of Rho. Similar interpretations can be applied to explain the results in Ref.~\citenum{perez2023simulating}. These insights underscore that the relative steady-state molecular populations can be straightforwardly deduced from the interplay between the linear polariton transmission spectra and the bare molecular absorption lineshape in the $N\to\infty$ limit.
 
 Finally, we note in passing that a similar explanation also provides some context for the celebrated result of Houdr{\'e}~\cite{houdre1996vacuum} on why for sufficiently strong light-matter coupling, polaritons inherit the homogeneous (Lorentzian) but not the inhomogeneous (Gaussian) linewidths of their parent molecules. In brief, far away from the center of the distribution, the height of the Gaussian tails is much smaller that for a Lorentzian with the same integrated area. This means that for large Rabi splittings, polariton transmission peaks overlap more significantly with the homogeneously broadened Lorentzian molecular tails. Correspondingly, the polariton peaks have a stronger inheritance of the Lorentzian linewidth. Furthermore, the overlap between the polariton modes and the molecular absorption tails is a key factor in determining how lossy these modes are, \textit{i.e.}, the more molecular absorption at the polariton frequencies, the lossier the mode, and fatter in frequency space, it becomes.

\subsection{Changes in photoreactivity via polaron decoupling }\label{sec:ex2}
\begin{figure}[!ht]
\centering
\includegraphics[width=.7\linewidth]{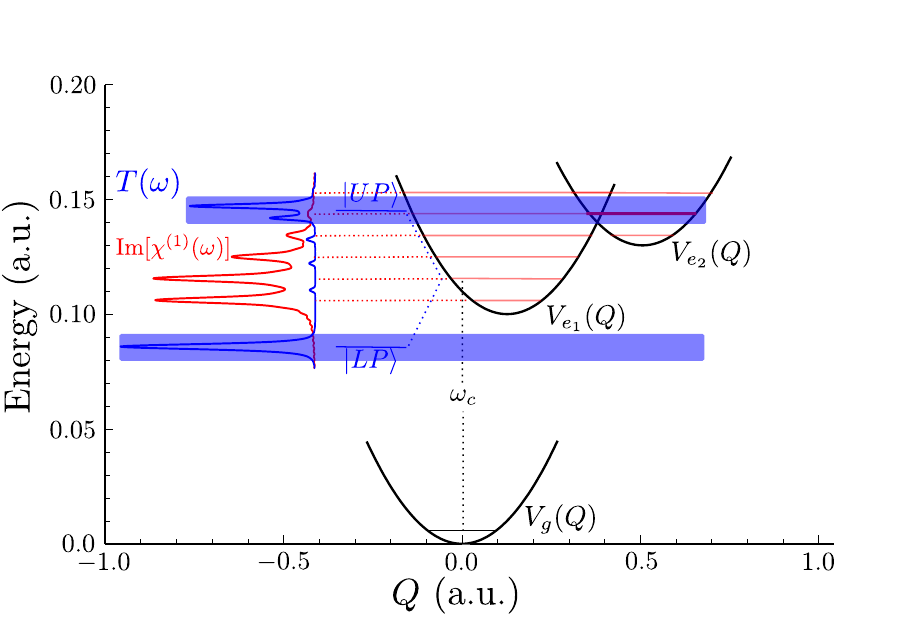}
\caption{\textit{Photoreactivity of molecular polaritons I.--} The molecule consider a has a single ground electronic state and two excited electronic states where only one of the ground-to-excited state transitions is bright.  The inset illustrates the spectral overlap, for zero disorder, between the linear polariton transmission $T(\omega)$ under strong coupling (blue) and the bare molecular absorption $\textrm{Im}[\chi^{(1)}(\omega)]$ outside of the cavity (red).  \label{fig:PESs}}
\end{figure}

\begin{figure*}[!ht]
\centering
\includegraphics[scale=.7]{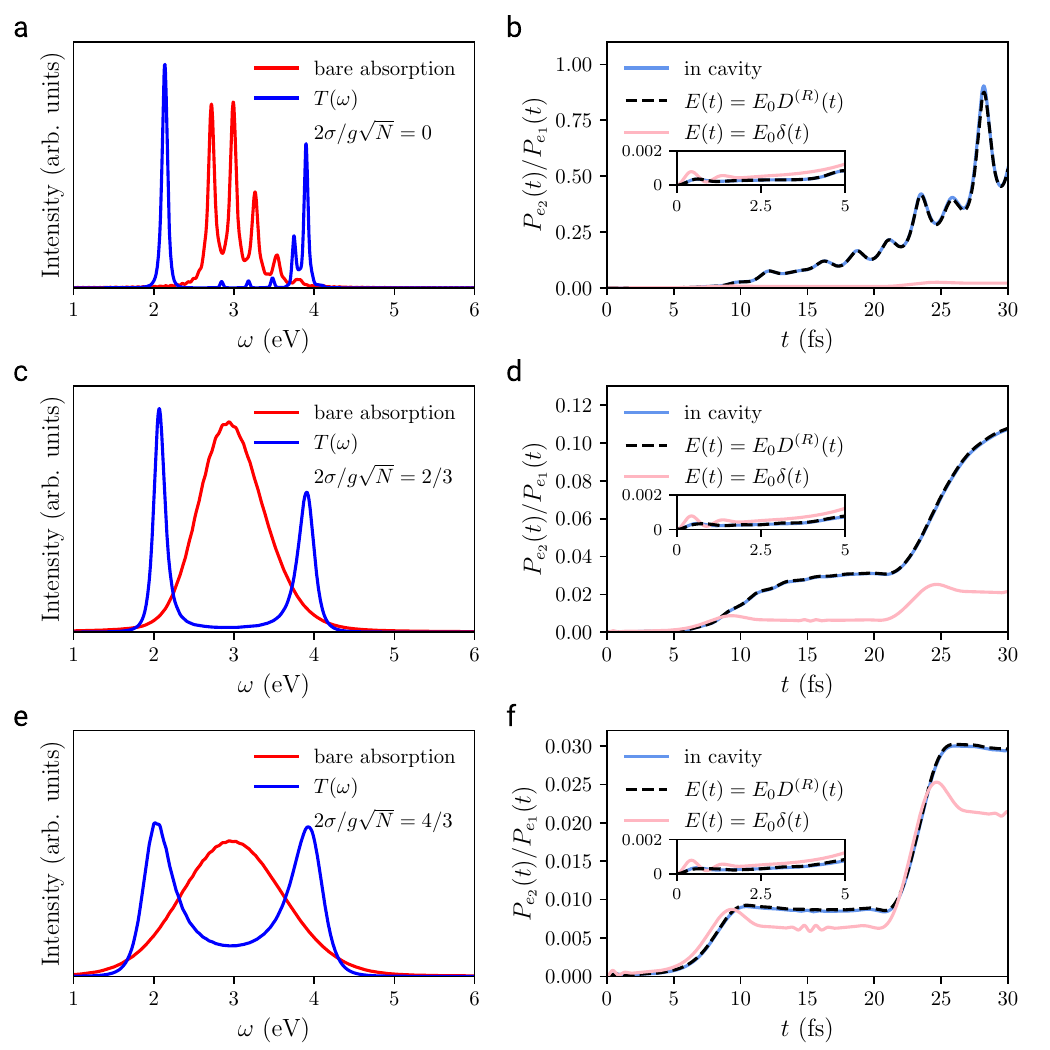}
\caption{\textit{Photoreactivity of molecular polaritons II.--} Panels (a,c,e) depict the overlap between the bare molecular absorption spectrum (red) with the polariton transmission spectrum $T(\omega)$ under strong coupling and a laser field with an identical intensity $|E(\omega)|^2$ (blue) for varying degrees of molecular disorder $\sigma$. The intracavity simulations were carried out with a collective light-matter coupling strength of  $g\sqrt{N}=0.816 \; \textrm{eV}$. Panels (b,d,f) present the corresponding short-time excited state dynamics both inside and outside the cavity. Outside the cavity, both a shaped pulse $E(t)=E_0D^{(R)}(t)$ and a broadband delta pulse $E(t)=E_0\delta(t)$ are considered. Notably, the relative population $P_{e_2}(t)/P_{e_1}(t)$ of state $e_2$ with respect to $e_1$ appears enhanced in the cavity when compared to the broadband pulse scenario outside of the cavity; however, the dynamics align exactly with those observed using an external pulse shaped to match the polariton transmission spectrum's intensity profile. As molecular disorder increases, the dynamics for both intracavity and shaped-pulse extracavity scenarios converge with those observed using a broadband pulse outside of a cavity. This convergence is attributed to the increased spectral overlap between $T(\omega)$ and $|E(\omega)|^2$ with the heterogeneously-broadened bare molecular absorption, mirroring the conditions of employing a broadband pulse externally. Note that at early times ($t<5$ fs), as highlighted by the insets, that the molecular disorder has little to no effect. \label{fig:short_time_dynamics}}
\end{figure*}

Here, we address the photoreactivity of a disordered ensemble of molecules coupled to a single cavity mode, utilizing a d-CUT-E effective Hamiltonian that was deployed previously~\cite{perez2023frequency} (Fig.\ \ref{fig:PESs}). A minimalistic molecular model is considered with a ground electronic state $g$ and two excited electronic states $e_1$ and $e_2$, where only the $g \to e_1$ transition is coupled to the cavity mode. Note that due to diabatic coupling between the two excited electronic states, the $g\to e_2$ transition borrows some oscillator strength from the $g\to e_1$ transition. In this model system, photoreactivity refers to the $e_{1}\to e_{2}$ transition probability, and is computed by the time-dependent population ratio $P_{e_{2}}(t)/P_{e_{1}}(t)$. The potential energy surfaces $V_{g/e_{1,2}}(Q)$ are parameterized by a single nuclear vibrational coordinate $Q$. The collective light-matter coupling is chosen such that the upper polariton (UP) transmission peak overlaps with states high in $e_2$ character; meanwhile, the lower polariton (LP) peak does not overlap with any significant features of the bare molecular spectrum (see inset of Fig.\ \ref{fig:PESs}). As in Section~\ref{sec:ex1}, we initially excite the system with a broadband pulse [\textit{i.e.}, a delta pulse in time $E(t)=E_{0}\delta(t)$] \textit{for both the inside and outside the cavity scenarios} (Fig.\ \ref{fig:short_time_dynamics}). The value of the constant field amplitude $E_{0}$ is chosen to be low enough to ensure we remain in the linear regime. Note that Eq. \ref{eq:absorption_reformulation} implies that, for steady state properties, a more meaningful comparison is made when the out-of-cavity scenario involves a weak laser with an intensity profile tailored to replicate the polariton transmission spectrum. The shaped (or ``filtered'') pulse acting on the bare molecules is one where the time-dependent electric field matches the strong coupling photon-photon correlation function, \textit{i.e.}, $E(t)=E_{0}D^{(R)}(t)$. We will focus on this scenario in Sec. \ref{sec:stcd}.

In the absence of disorder (Fig.\ \ref{fig:short_time_dynamics}b), initial observations suggest that strong coupling significantly alters photoreactivity compared to the bare molecule scenario, as evidenced by an increased $e_2$ to $e_1$ population ratio $P_{e_2}(t)/P_{e_1}(t)$ at time $t\sim 30$ fs. As molecular disorder increases (Figs.\ \ref{fig:short_time_dynamics}d and \ref{fig:short_time_dynamics}f), the temporal behavior of this population ratio gradually aligns with that observed outside the cavity.

We now draw from the insights gained in Section~\ref{sec:ex1}. In the disorder-free scenario, the upper polariton (UP) transmission window coincides with a peak in the bare molecular spectrum predominantly characterized by $e_2$, while the lower polariton (LP) transmission overlaps slightly with the low-energy tail of the molecular spectrum (see Fig.\ \ref{fig:PESs} inset). Eq.\ ~\ref{eq:absorption_reformulation} aids in rationalizing the observed photoreactivity enhancement under strong coupling: the light that the cavity selectively channels through the UP transmission window overlaps with the portion of the bare molecular absorption spectrum that is rich in $e_2$ product character. As disorder increases, the polariton transmission spectrum more closely aligns with the reactant features of the bare molecular spectrum (see Figs. \ref{fig:short_time_dynamics}c and \ref{fig:short_time_dynamics}e), leading to a scenario where the linear absorption of broadband light within the cavity converges towards how a broadband pulse would be absorbed in free space. This explains why, with increasing molecular disorder, our intracavity simulations tend towards the out-of-cavity scenario under broadband excitation.

These results are in alignment with recent reports in the literature. Thomas \textit{et al.}~\cite{thomas2023nonpolaritonic} have reported experimental modification of the photoisomerization of spiropyran to merocyanin under strong coupling, proposing that the change was due to a non-polaritonic effect predominantly influenced by molecular absorption of ultraviolet radiation within the cavity. Concurrently, Dutta \textit{et al.}~\cite{dutta2024thermal} explored the photochemistry of 10-hydroxybenzo[h]quinoline under strong coupling through molecular dynamics simulations and experiments, suggesting a key role played by the congruence between bright polaritons and molecular dark states. As a result, the photoreactivity changes under strong coupling because the available light within the cavity has a distinct frequency distribution from the optical pump outside the cavity. For instance, if the polaritonic transmission windows coincide with the bare molecular energy levels that lead to a desired reaction pathway, that pathway will be selectively enhanced due to the increased \textit{relative} absorption at those frequencies. Conversely, if the polaritonic windows overlap with unreactive bare molecular states -- or no molecular states -- the photoreactivity can be suppressed. 

This linear optics viewpoint also provides an explanation for the strong coupling control of photochemical processes through polaron decoupling~\cite{herrera2016cavity,galego2016suppressing}, albeit, in the $N\to \infty$ and low temperature limits. This phenomenon has been explained as a version of motional narrowing where the exchange of energy between light and matter is so fast that the LP does not ``feel'' much coupling to the molecular vibrations, with the consequence of suppressed reactivity. Calculations show that the UP does not enjoy such polaron decoupling to the same degree, presumably due to strong mixing with high-lying vibronic states~\cite{zeb2018exact}. Similar to the inset of Fig. \ref{fig:PESs}, typically under polaron decoupling the LP transmission peak is off-resonant from any bare molecular vibronic transitions, so little to no absorption will occur upon pumping the LP. Therefore, we expect that no nuclear motion will occur upon excitation of the LP, leading to a corresponding decrease in photoreactivity~\cite{herrera2016cavity,galego2016suppressing}. Again, there is nothing inherently incorrect about attributing the suppression of reactivity to polaron decoupling of the LP potential energy surface, but we believe it is much simpler to understand it in terms of the lack of molecular absorption at the LP frequency. It is important to note that Ref.~\citenum{herrera2016cavity} also considers finite temperature effects, which go beyond the approximations of CUT-E and d-CUT-E.  

Importantly, and regardless of interpretation, the same suppression or enhancement in the low temperature limit can be achieved outside of a cavity by driving at the corresponding polariton frequencies: pumping free-space molecules at the LP frequency corresponds to off-resonantly pumping the system, so little photoreactivity is expected, while pumping at the UP frequency corresponds to resonantly driving the product peak so an increase in photoreactivity \textit{is} expected. This selective enhancement or suppression of reaction pathways based on the overlap of polaritonic and bare molecular states underscores the nuanced interplay between polaritonic effects and bare molecular linear absorption. Again, this review highlights the importance of considering these effects in the analysis of the excited state dynamics of polaritonic systems and the importance of providing the correct comparisons for intracavity and free-space scenarios. It is crucial to emphasize that the linear polaritonic response for systems under strong coupling with sufficiently small $N$ are not accurately captured by Eq.\ \ref{eq:absorption_reformulation}. Consequently, changes in photoreactivity under strong coupling in such cases \textit{cannot} be described using this optical filtering viewpoint.

\subsection{Coherent dynamics in the limit of many molecules}\label{sec:stcd}

\begin{figure}[!ht]
\centering
\includegraphics[scale=1.2]{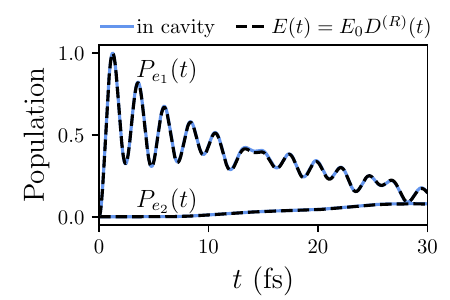}
\caption{\textit{Coherent dynamics in the large-$N$ limit.--} Population dynamics of the first, $P_{e_1}(t)$,  and second, $P_{e_2}(t)$, excited state populations inside the cavity versus when the system outside of the cavity is driven with a time-dependent pulse $E(t)=E_{0}D^{(R)}(t)$ that has the same intensity profile as the polariton transmission spectrum. Note that the populations are normalized to the maximum value of $P_{e_1}(t)$. The dynamics, up to a constant factor that depends on the cavity-enhancement of the incoming electromagnetic field, are identical, with both scenarios presenting Rabi oscillations in the $e_1$ population dynamics, while no such observations are observed in the $e_2$ population dynamics since this state is not directly coupled to the cavity. This calculation shows the importance of making the correct in- and out-of-cavity calculations and experiments. \label{fig:rabi oscillations}}
\end{figure}

Using the same model introduced in Section \ref{sec:ex2}, we now present an in-depth comparison between the excited state dynamics of a polaritonic system under broadband excitation with $E(t)=E_{0}\delta(t)$ and that of the bare molecular ensemble pumped with the filtered pulse $E(t)=E_{0}D^{(R)}(t)$. Simulations using the CUT-E method confirm that the photoreactivity of the bare molecules triggered by $E(t)$ is, up to a constant factor, \textit{identical} to that observed inside the cavity (see Figs. \ref{fig:short_time_dynamics}b, \ref{fig:short_time_dynamics}d, \ref{fig:short_time_dynamics}f and Fig.\ \ref{fig:rabi oscillations}). 

\begin{figure*}[!ht]
\begin{centering}
\includegraphics[width=\linewidth]{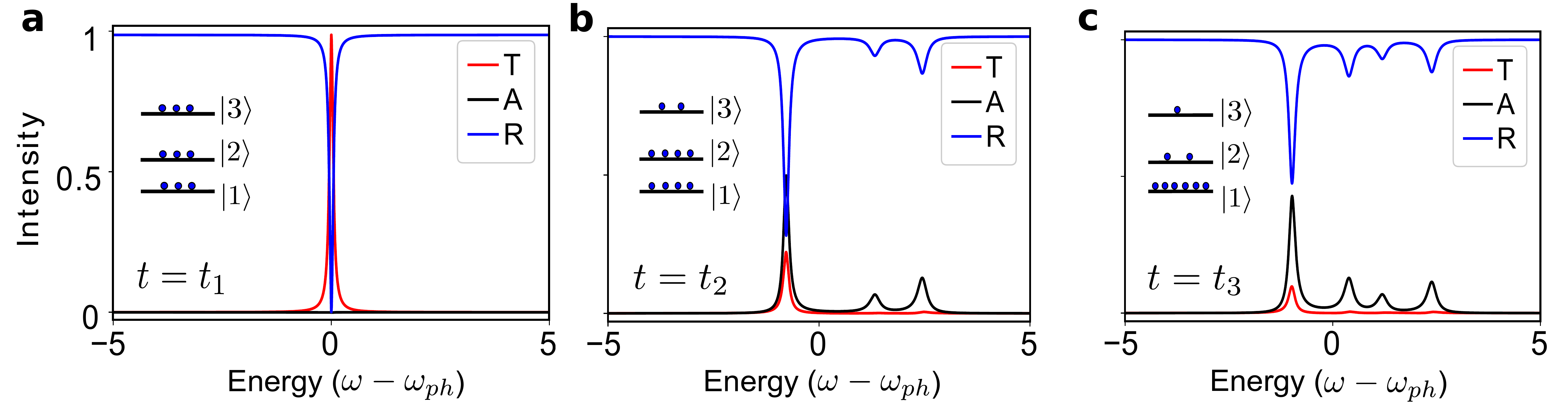}
\end{centering}
\caption{\textit{Incoherent nonlinear spectroscopy of polaritons.--} A representative example of a polaritonic incoherent nonlinear spectroscopic experiment showing transmission, reflection, and absorption for an ensemble of identical three-level systems at different delay times, $t$ such that $t_{i+1}-t_{i}\gg\tau$ where $\tau$ is the dephasing timescale of the system. The three states are labeled as $|1\rangle$, $|2\rangle$, and $
3\rangle$ and the total transition dipole moment of the three-level system is $\mu=\mu_{12}+\mu_{23}+\mu_{13}$, where ${\mu_{ij}}=\mu_{ij}|j\rangle\langle i|+\mu_{ji}|i\rangle\langle j|$ for $i,j\in\{1,2,3\}$. The configurations of the system, obtained upon optical pumping, are shown in the insets. Configuration (a) represents the state created upon optical pumping with complete saturation, rendering the medium transparent to the cavity. After a time of $t_2-t_1$, the system relaxes to (b) where the $\mu_{12}$ transition is still saturated, giving us three polariton peaks. The system then further relaxes to reach (c) at time $t_3$, and we can see all four polariton peaks appearing owing to all the molecular transitions, $\mu=\mu_{12}+\mu_{23}+\mu_{13}$, being active. The time-varying total (linear +nonlinear) spectra have been computed using Eq.\ ~\ref{spectra_formula} with an effective time-dependent molecular susceptibility, $\chi^{(1)}(\omega;t)$, that adjusts adiabatically to each configuration. (Reproduced from Ref. \citenum{yuen2023linear}, Copyright 2024 J. Yuen-Zhou and A. Koner, published by AIP Publishing.)\label{three_LS}} 
\end{figure*}

These results contradict the common misconception that collective strong light-matter coupling, evidenced by time-domain Rabi oscillations or a frequency-domain Rabi splitting in the linear polaritonic response, entails a local hybridization of light and matter that fundamentally alters the structure at the single-molecule level, a phenomenon that requires \textit{ab-initio} quantum mechanical treatments for its description \cite{Ruggenthaler2023understanding,Sidler2024unraveling,Horak2025analytic}. Here we show that the molecular dynamics under collective strong light-matter coupling and upon broadband excitation can be reproduced outside of a cavity via constructive and destructive interference between excited state amplitudes promoted at different times by a filtered pulse whose field is enhanced by a factor equal to the cavity quality factor, (Fig.\ \ref{fig:rabi oscillations}). In other words, the bare molecules are illuminated with light at the two frequencies resonant with the upper and lower polaritons, giving rise to beating of the transmitted intensity at the Rabi frequency, which decays as light is lost to molecular absorption and cavity leakage. This phenomenon is referred to as linear wavepacket interferometry in the field of ultrafast spectroscopy~\cite{scherer1991fluorescence,tannor2007introduction}. Building upon this concept, nonlinear wavepacket interferometry has been proposed as a means to achieve genuine ``pump-dump'' control and detection of molecular dynamics outside the FC region~\cite{BrumerShapiro,TannorRice}. It is important to clarify that the ultrafast polaritonic dynamics will only be reproduced by bare molecules excited outside of cavity if the optically filtered source is coherent and has a frequency-dependent phase that matches the one for $D^{(R)}(\omega)$~\cite{Brumer}. In the scenario where the optically filtered pulse matches the polariton transmission spectrum only in intensity profile, the mismatch in phase information can cause the transient dynamics for the molecules outside of cavity to differ from the dynamics under strong coupling before both systems reach the same steady-state.  Similar considerations have been discussed in the context of one-photon phase control experiments.~\cite{Brumer2010conditions,Arango2013onephoton}. Importantly, if the $1/N$ effects (\textit{e.g.}, dark-state-to-polariton relaxation) become significant before the systems reach stead-state, the dynamics under strong coupling versus under linear excitation via a filtered pulse \textit{will not be the same} (see Sec. \ref{sec:cute}). Finally, it could be argued that polaritons can facilitate one-photon phase control without the need of external pulse shapers. However, current state-of-the-art technologies routinely achieve resolutions of the order of 0.001 nm~\cite{Weiner2011ultrafast}, much higher than those of typical molecular polaritons (commonly in the order of tens of nanometers at room temperature in the condensed phase). Nonetheless, overcoming the resolution limitations could be a new research direction in the molecular polaritons field.

The results in Fig. \ref{fig:rabi oscillations} indicate that choosing the same initial conditions inside and outside of the cavity when performing quantum dynamics simulations can be misleading. Initial conditions outside the cavity must be sampled in a way that  accounts for the filtering of the hypothetical laser amplitude (\textit{not just intensity!}) acting on the polaritonic system.

\subsection{Incoherent pump-probe spectroscopy of polaritons}
As discussed in Section~\ref{sec:impurity_model}, the linear response implications of the quantum impurity model for $N\rightarrow\infty$ hold for arbitrary initial states so long as the quantum states of light and matter are decoupled, \textit{i.e.}, $\rho=\rho_{ph}\otimes\rho_{mol}$, and are stationary with respect to $H_0$. These conditions hold true not only for thermal states but also for nonequilibrium stationary states resulting from optical pumping. As we shall argue, the stationarity condition can be liberally taken to mean that the molecular dynamics are slow compared to the inverse of the spectral resolution. This condition is particularly relevant for incoherent nonlinear spectroscopy experiments~\cite{dunkelberger2016modified,wei2018nonlinear1,delpo2021polariton,renken2021untargeted,cheng2022molecular,hirschmann2023role}, where, upon optical pumping, polaritons relax into incoherent dark states but do not fully thermalize, provided that the timescale of the dark state dynamics $\tau\gg \frac{2\pi}{\Gamma}$, where $\Gamma$ is the minimum of the cavity and molecular dephasing linewidth. $\Gamma$ characterizes the decay rate of the photon-photon correlation function and hence dictates the resolution of the acquired spectrum. Thus, any dynamics slower than the corresponding timescale $\tau$ can be treated adiabatically, ensuring the validity of the stationarity condition within the spectral resolution of the polaritonic spectrum. In other words, the molecular state can be regarded as frozen on the timescale of $\tau$. The response triggered by a probe pulse can be regarded as the linear response of this ``stationary state'' which changes parametrically as a function of a coarse-grained waiting time $\tau$ [sampled at intervals coarser than $2\pi/\Gamma$] and can be characterized by an effective (time-dependent) linear molecular susceptibility $\chi^{(1)}(\omega;t=n\tau)$. This picture is completely analogous to what happens in incoherent pump-probe experiments outside of a cavity.~\cite{pascal2017incoherent,paff1989nonlinear,Ashworth1996vibronic}

Fig.~\ref{three_LS} presents an illustrative model for a hypothetical incoherent pump-probe experiment, showing the transmission, reflection, and absorption for an ensemble of identical three-level systems after optical pumping. Panels a, b, and c of Fig.\ ~\ref{three_LS} represent the total (linear+nonlinear) signal of the system at three stationary non-thermal configurations after delay times $t=t_1<t_2<t_3$ respectively. In the regime where $\Delta t=t_{i+1}-t_i$ is greater than $\tau$,  Eq.~\ref{spectra_formula} can be used to compute the spectroscopic observables. The configurations of the system, obtained upon optical pumping, are (a) $p_{1}=p_{2}=p_{3}$, (b) $p_{1}=0.45$, $p_{2}=0.45$, $p_{3}=0.1$, and (c) $p_{1}=0.7$, $p_{2}=0.2$, $p_{3}=0.1$, at delay times $t=t_1<t_2<t_3$ respectively. The parameters used are: $\omega_{12}=\omega_{ph}=1, \omega_{23}=2\omega_{ph}$, $\kappa=0.1, \gamma=0.3$, and the collective light-matter coupling $g\sqrt{N}=1$ (arbitrary frequency units). Configuration (a) represents the state created upon optical pumping with complete saturation, rendering the medium transparent to the cavity. Thus, we see the bare cavity spectra in the linear response. After a time of $t_2-t_1$, the system relaxes to (b) where the $|1\rangle\rightarrow|2\rangle$ transition is still saturated and transparent. Hence, we see three polariton peaks owing to two molecular transitions, $\mu=\mu_{13}+\mu_{23}$, coupling off-resonantly to the cavity. This configuration is still in a non-thermal stationary state. The system further relaxes to reach (c) at time $t_3$ where all three molecular transitions are active and couple to the cavity; $|1\rangle\rightarrow|2\rangle$ resonantly, and $|2\rangle\rightarrow|3\rangle$ and $|1\rangle\rightarrow|3\rangle$ off-resonantly -- leading to the appearance of four polariton peaks. 

Thus, the implications of the optical filtering presented in the previous sections hold, but with an important caveat:  the polariton windows change in time, adiabatically adjusting to $\chi^{(1)}(\omega;t)$ at each configuration. It is therefore not surprising that many experimental ultrafast nonlinear spectroscopy studies~\cite{dunkelberger2016modified,wei2018nonlinear1,delpo2021polariton,renken2021untargeted,cheng2022molecular,hirschmann2023role,pyles2024Revisiting} successfully use transfer matrix methods with an effective time-dependent linear molecular susceptibility to model transient polaritonic transmission spectra. 

\subsection{\label{sec:rp}Radiative pumping as described by linear optics}

\begin{figure*}[!ht]
\centering
\includegraphics[width=\linewidth]{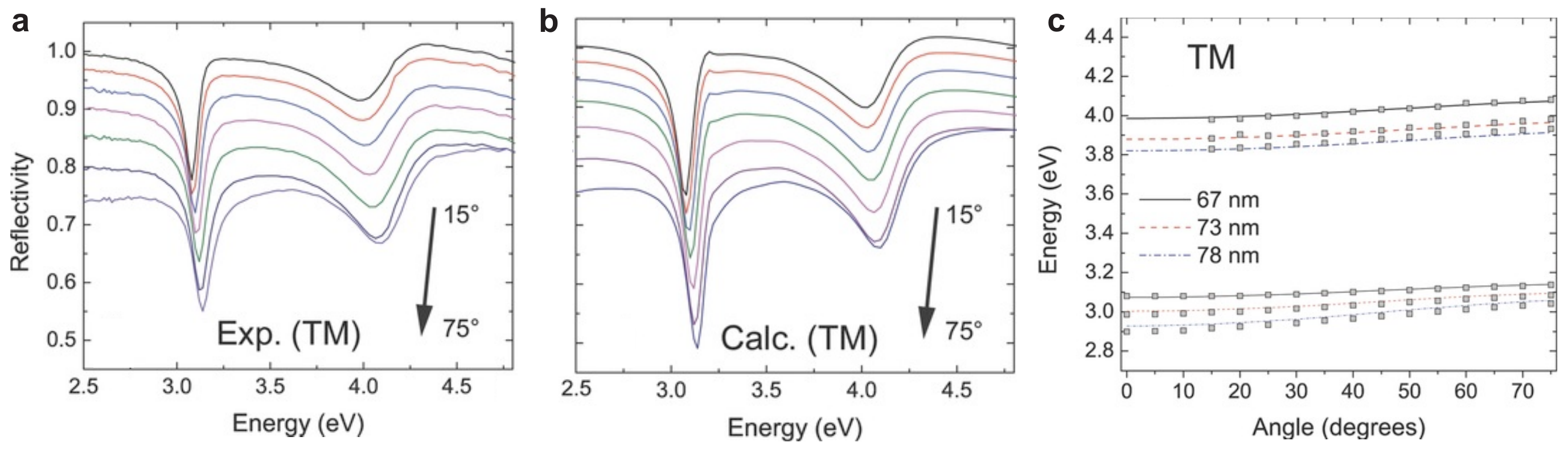}
\caption{\textit{Classical optics modeling of polariton observables.--} The angle-dependent transverse magnetic (TM)-polarized reflectivity spectra measured for a metal-organic-metal microcavity (a) are accurately reproduced using the transfer matrix method (b). The polariton dispersion in both cases is consistent with a full Hopfield Hamiltonian treatment in the ultrastrong coupling regime for several different cavity thicknesses (c). (Reproduced from Ref. \citenum{KenaCohen2013ultrastrongly} with permission from Wiley-VCH, Copyright, 2013.) \label{KenaRefSims}} 
\end{figure*}

Relaxation mechanisms from dark states back to polariton states are $1/N$ effects of great interest in the molecular polaritonics community, as they play important roles in applications such as polariton-assisted remote energy transfer~\cite{paret,Feist2020ET}, polariton transport~\cite{Schachenmayer2015transport,Mukherjee2018,Schwartz2018transport,suyabatmaz2023vibrational,Delor2023transport,Groenhof2024transport}, and polariton condensation~\cite{Yamamoto2010EPBEC,kena2010room,Keeling2011EPC,daskalakis2014nonlinear,Yamamoto2014EPC,Keeling2017NewEra,KenaCohen2017superfluidity,Sindhana2022condensate,Vinod2024Direct}. 
The most frequently observed relaxation mechanism is radiative pumping, which consists of spontaneous emission from incoherent excitons into the polariton modes. This picture was first drawn in seminal works by Michetti and La Rocca, who provided a phenomenological rate for radiative pumping that is proportional to the bare molecular emission rate at the frequency of the polariton mode, multiplied by the photonic Hopfield coefficient of the hybrid mode.~\cite{michetti2008simulation,michetti2009exciton}. This description is consistent with experimental observations~\cite{coles2014polariton,grant2016efficient,Keeling2020BEC,Tomo2022Low,Jiang2022exciton,Prathmesh2023radiative,Castellanos2023nonequilibrium}.

This rate can be rigorously derived from the Hamiltonian introduced in Eq. \ref{eq:Hamiltonian}, as the total radiative decay of the so-called Stokes-shifted state (an incoherent exciton) using first-order perturbation theory in the single-molecule light-matter coupling $g$ \cite{Perez2025radiative}. The resulting radiative pumping rate written in terms of the linear optical properties in Eqs. \ref{eq:absorption}-\ref{eq:reflection} yields\cite{Perez2025radiative},
\begin{equation}\label{eq:rp}
    \Gamma_{rp}(\omega)=\left(\frac{g^{2}}{\kappa_{L}}\right)\sigma_{em}(\omega) \left[A(\omega)+\left(\frac{\kappa}{\kappa_{R}}\right)T(\omega)\right].
\end{equation}
This shows that the frequency dependent rate $\Gamma_{rp}(\omega)$ of radiative pumping is proportional to the product of the free-space molecular emission $\sigma_{em}(\omega)$ with the polariton absorption and transmission spectra. Eq. \ref{eq:rp} separates the rate into two components: one proportional to $A(\omega)$ indicating re-absorption of emitted light by the polaritons, and one proportional to $T(\omega)$ indicating transmission through the cavity mirrors. Moreover, the prefactor $g^{2}/\kappa_{L}\propto \mathcal{Q}/\mathcal{V}_{ph}$ encodes cavity enhancement of the molecular emission. Ignoring the reabsorbed component, photoluminescence via radiative pumping corresponds to the enhanced molecular emission filtered through the polariton transmission spectrum. Although this is intuitive, we must keep in mind that $A(\omega)$ and $T(\omega)$ are defined for an out-of-cavity, rather than an inside-of-cavity, excitation source (see Fig. \ref{fig:process}a for an example of the former).

From a classical optics perspective, radiative pumping can be modeled by an oscillating dipole that radiates energy into the polariton modes (\textit{i.e.}, emission of a photon at the particular internal angle(s) corresponding to a polariton mode), which can be described by standard dipole emission models \cite{Benisty1998method,Sullivan1997enhancement,Neyts1998simulation}. In this context, the radiative pumping rate corresponds to the dipole power dissipated into the range of in-plane wavevectors that correspond to the polariton mode at each frequency in the free-space spontaneous emission spectrum.

In Sections \ref{sec:polrec} and \ref{sec:vr} we discuss the effects of the reabsorbed component of the radiative pumping and higher-order processes in the single-molecule light-matter coupling $g$, that render this linear optics description of polariton photoluminescence incomplete.

\subsection{Experimental examples of optical filtering}
Insofar as polariton properties such as Rabi splitting, linewidth, and dispersion are generally well described by classical optics using the dielectric functions of the bare materials comprising the system (Fig.~\ref{KenaRefSims}), all linear optical properties of polaritons -- from the strong to ultrastrong coupling regime~\cite{KenaCohen2013ultrastrongly} -- can be interpreted within the context of optical filtering. There is less clarity on the extent to which the nonlinear optical properties of polaritons also originate from optical filtering, defined here as a nonlinear response of the cavity system that can be understood using classical optics based on the nonlinear response of the bare material outside the cavity.

Initial work on polariton-enhanced second harmonic generation (SHG)~\cite{Ebbesen2016shg} and nonlinear absorption/refraction~\cite{Ebbesen2021large} concluded that the enhancement exceeded what was expected from optical filtering, while subsequent work on third harmonic generation (THG) concluded the opposite~\cite{Barachati2018thg}. That is, the increased THG output of the cavity sample was well explained when taking into account the cavity-modified field strength at the fundamental and third harmonic frequencies interacting with the $\chi^{(3)}(3\omega,\omega,\omega,\omega)$ susceptibility of the bare material. The case of electroabsorption (another $\chi^{(3)}$ nonlinearity) was also found to be a filtering effect and highlighted differences in language that ultimately describe the same physical phenomenon.~\cite{Cheng2022electroabsorption} From the polariton perspective, polaritons exhibit an enhanced Stark shift in response to a DC electric field whenever the polariton modes become resonant with other (weakly coupled) states in the system, as expected from standard perturbation theory. On the other hand, the same effect viewed from the classical optics perspective originates from the field-induced change in energy and oscillator strength of the underlying material transitions, the visibility of which is resonantly enhanced when observed through a mode of the system (\textit{i.e.}, a polariton). These views are two sides of the same coin and are related by a change of basis.
\begin{figure*}[!ht]
\centering
\includegraphics[width=.8\linewidth]{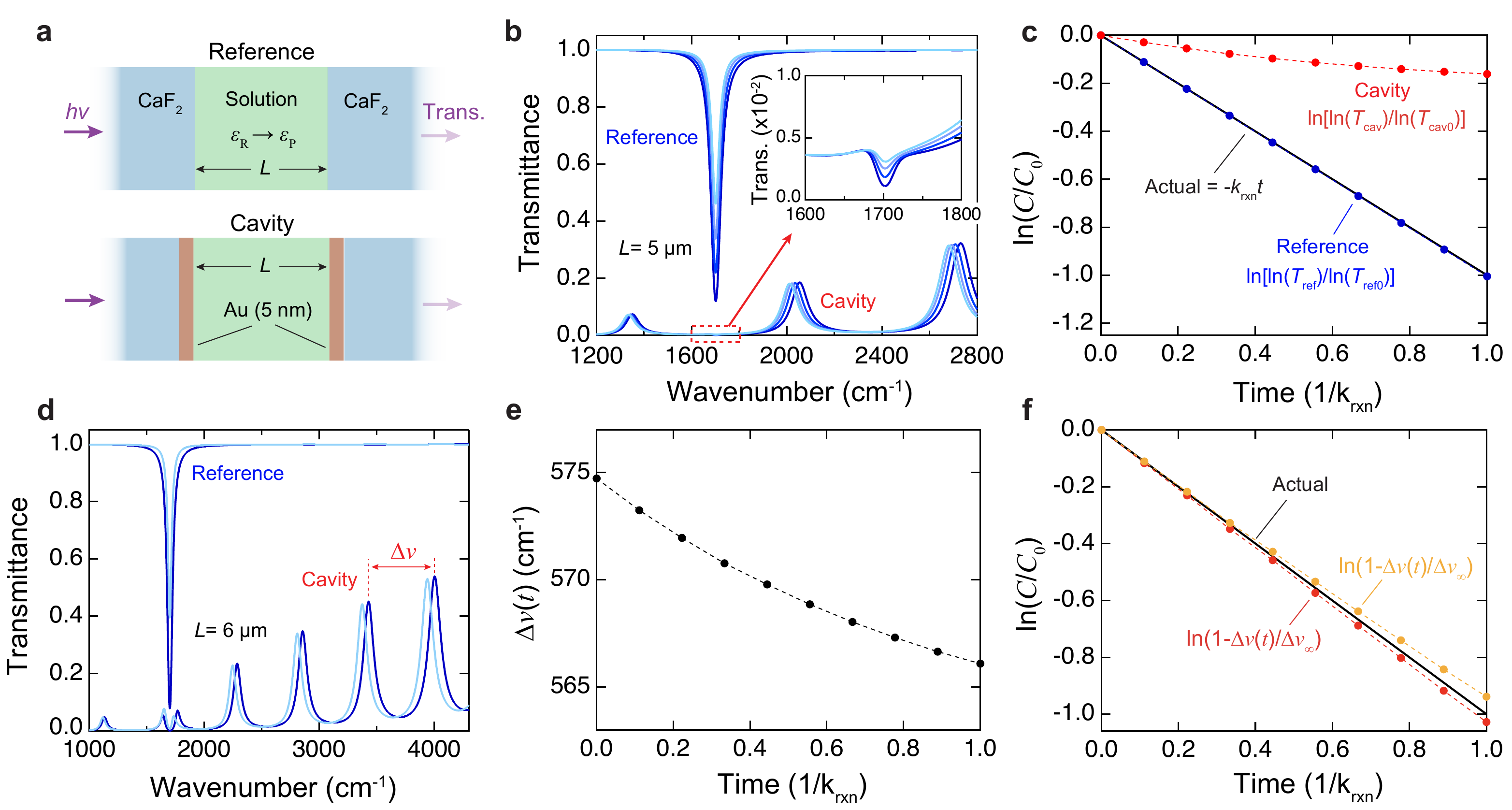}
\caption{\textit{Polariton pseudochemistry from microcavity optics.--} (a) Schematic of a simple transfer matrix model for reference and cavity sample cells filled with a solution undergoing a first order reaction from reactant ($R$) to product ($P$) with rate constant, $k_\mathrm{rxn}$, according to the scheme \ce{$R$ ->[$k_\mathrm{rxn}$] $P$}. The reactant has a dielectric function, $\varepsilon_{R}=\varepsilon_{R\infty}+\omega_0^2 f/(\omega_0^2-\omega^2-i\omega\gamma)$, that reflects a vibrational transition at $\omega_0=1700$~cm\textsuperscript{-1} with damping rate $\gamma=30$~cm\textsuperscript{-1}. The product does not have this vibrational transition and has a different high frequency dielectric constant, $\varepsilon_\mathrm{P}=\varepsilon_{P\infty}$. The time-dependent dielectric function is therefore $\varepsilon (t)=\varepsilon_{R}e^{-k_\mathrm{rxn}t}+\varepsilon_\mathrm{P}(1-e^{-k_\mathrm{rxn}t})$. This example assumes $\varepsilon_{R\infty}=2$ and $\varepsilon_{P\infty}=2.1$, similar to experiments\cite{thomas2016ground,thomas2019tilting}. (b) Time-dependent transmittance spectra (at normal incidence; lighter blue corresponds to later time) for the reference and cavity simulations when $f=10^{-2}$, in a scenario where the cavity is detuned from the vibrational mode, which is just used as a spectator to track consumption of the reactant as in Ref.~\citenum{hiura2021}. (c) Relative reactant concentration inferred from the transmittance of the reference and cavity samples at $\omega_0$. The rate (equal to the slope) inferred for the cavity case is different than the actual rate due to the impact of the changing cavity phase that accompanies the loss of the vibrational transition\cite{imperatore2021reproducibility}. (d) The same model, except with the cavity length adjusted to $L=6$~µm to achieve VSC. (e) Time-dependent change of the free spectral range, $\Delta\nu$, of the higher order cavity modes indicated in (d). \textbf{(f)} Relative reactant concentration inferred inside the cavity using the relation, $\ln{(1-\Delta\nu(t)/\Delta\nu_\mathrm{\infty})}$, where $\Delta\nu_\infty$ is the change in free spectral range at $t\rightarrow\infty$ \cite{thomas2016ground,thomas2019tilting}. Owing to the dispersive refractive index, $\Delta\nu_\infty$ is not constant, and so the apparent rate changes depending on what frequency range $\Delta\nu_\infty$ is determined from. For example, if $\Delta\nu_\infty$ is computed from the same higher order modes that the reaction rate measurement is based on in (d,e), then this approach slightly overestimates the reaction rate (red line). However, if $\Delta\nu_\infty$ is based only on the background refractive indices of the reactant and product at high frequency (\textit{i.e.}, $\Delta\nu_\infty=(2L)^{-1}(\varepsilon_{R\infty}^{-1/2}-\varepsilon_{P\infty}^{-1/2})$ in the high frequency limit), then this approach slightly underestimates the rate (orange line).\label{pseudochem}}
\end{figure*}

Within the context of chemistry, optical filtering effects can distort the reaction rate that is inferred from time-dependent changes in transmittance relative to a non-cavity control (Fig.~\ref{pseudochem}a,b,c). The difficulty stems from the fact that, outside of a cavity, changes in the real part of the refractive index (\textit{i.e.}, phase) have a negligible effect on sample transmittance, whereas inside of a cavity, changes in index have a significant effect since they alter interference within the cavity. Because a change in absorbance of a reactant or product species (which would normally be used to infer concentration) necessarily leads to a change in index due to the Kramers-Kronig relation, the reaction rate inferred from cavity transmittance appears different than that of a non-cavity control, even when the underlying reaction rate is the same in both cases as shown in Fig.~\ref{pseudochem}(c)\cite{imperatore2021reproducibility}. The practice of inferring reaction rate from the time-dependent shift of higher order cavity modes\cite{thomas2016ground,thomas2019tilting,thomas2019nanophotonics,hirai2020} can also be problematic if not done carefully. As detailed in Ref.~\citenum{thomashigherorder2024}, the change in dielectric constant from an evolving strongly coupled transition at low frequency persists to a small degree even at high frequency and can skew the reaction rate that is inferred from the shift in cavity free spectral range of higher order cavity modes (Fig.~\ref{pseudochem}d,e,f). 

Recent work by Ahn \textit{et al.}~\cite{ahn2023modification} described a systematic fitting procedure to circumvent these types of problems and accurately determine the time-dependent reactant concentration from cavity transmittance measurements, providing strong evidence for genuine VSC suppression of the reaction rate. The most significant impact of optical filtering on polariton chemistry to date was reported by Thomas \textit{et al.}~\cite{thomas2023nonpolaritonic}, who concluded that the original cavity-enhanced photoisomerization reaction reported by Hutchison \textit{et al.}~\cite{hutchison2012modifying} could be explained by optical filtering based on cavity-induced changes in absorption of the excitation light. This finding contradicts work by Zeng \textit{et al.}~\cite{zeng2023control}, who reproduced the original results by Hutchison. Thomas' reinterpretation has also been challenged by Schwartz \textit{et al.}~\cite{schwartz2024importance}, who disputed the role of altered pump light absorption in the original experiments.

\section{Beyond optical filtering: Nontrivial polaritonic effects}\label{sec:beyond_filtering}

In Sec. \ref{examples}, we presented scenarios where polaritonic effects can be understood as optical filtering. In these examples we considered the large-$N$ limit, and, for the most part, neglected the ``$1/N$ effects'' that lead to processes like relaxation of dark states back to polaritons. In the $N\to\infty$ limit, the linear optics treatment is exact when considering the first excitation manifold. Consequently, a laser pulse shaped to mimic the polariton transmission spectrum can be employed outside of the cavity to achieve the same ``polaritonic'' phenomena. Even upon relaxation of the $N\to\infty$ limit, the dark-state-to-polariton relaxation rate via radiative pumping can be partially described as optical filtering of the cavity-enhanced fluorescence of incoherent excitons through the polariton windows. Still, there are many situations where polaritonic phenomena go beyond optical filtering. In the following sections, we present some $1/N$ effects that cannot be described as simply optical filtering: polariton-assisted photon recycling and vibrational relaxation. We then consider scenarios that challenge the limitations of the theoretical approach presented in Sec.~\ref{sec:impurity_model}, including when the number of excitations approaches the number of molecules per mode and the scenarios where few-molecule strong coupling can be achieved. Finally, we mention some experiments conducted in the collective strong coupling regime in the large-$N$ limit, yet where polaritonic phenomena are observed that cannot be explained via optical filtering. 

\subsection{\label{sec:polrec}Polariton-assisted photon recycling}
\begin{figure}[!ht]
\centering
\includegraphics[width=1\linewidth]{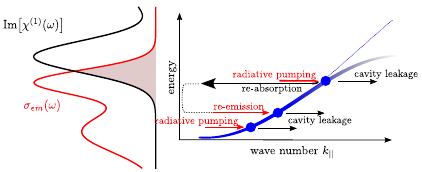}
\captionof{figure}{\textit{Mechanism of polariton-assisted photon recycling.--} A fraction of the light emitted from the dark states can be reabsorbed by the polaritons. This fraction is proportional to the bare molecular absorption Im$\left[\chi^{(1)}(\omega)\right]$ (generally high frequencies of the emission spectrum). Subsequent re-emission can cause photoluminescence intensity to differ from that predicted by filtering of the bare emission $\sigma_{em}(\omega)$ through the polariton transmission spectrum.\label{fig:recycling}}
\end{figure}

In Sec.~\ref{sec:rp} we describe radiative pumping as emission from incoherent excitons, and express its rate in terms of linear spectroscopic observables (see Eq. \ref{eq:rp}). We find that light emitted from dark states can not only be transmitted out of the cavity, but also be re-absorbed by the material. This corresponds to radiative pumping from dark to polariton states, and subsequent decay from polaritons back to dark states \cite{Perez2025radiative}. Since the light emitted by the dark states can be re-absorbed and then subsequently re-emitted (a second instance of decay from dark to polariton states), the total observed photoluminescence can not be described by the product of the bare molecular emission and the polariton transmission spectra.

Recent \textit{ab initio} quantum dynamics simulations showed that most of the polariton population decays to the dark states instead of via cavity leakage, especially for highly disordered ensembles.~\cite{sokolovskii2023multi} Here we confirm that this is a direct consequence of the polaritonic transmission spectrum overlapping significantly with the bare molecular absorption. We calculate the ratio between the re-absorption and transmission components of Eq. \ref{eq:rp} in the large-$N$ limit using Eqs. \ref{eq:absorption} and \ref{eq:transmission}. The resulting expression is equal to the product of the cavity quality factor and bare molecular absorption:
\begin{equation}    
\frac{A(\omega)}{\left(\frac{\kappa}{\kappa_{R}}\right)T(\omega)}= \mathcal{Q}\ \textrm{Im}\left[\chi^{(1)}(\omega)\right].
\end{equation}
Given the above, the process of photons being re-absorbed and re-emitted multiple times before leaving the cavity is enhanced in the collective strong coupling regime -- since Im$\left[\chi^{(1)}(\omega)\right]$ is roughly proportional to $g^2N$, see Ref.~\citenum{schwennicke2024extracting}. This is particularly evident when the bare molecular emission and absorption spectra overlap. In such cases, photoluminescence cannot be explained simply as optical filtering of the bare emission through the polariton windows. From a classical optics viewpoint, an iterative application of the dipole emission modeling procedure (similar to the case of resonant cavity light-emitting diodes) is needed to accurately describe this scenario~\cite{Benisty1998impact}.

This mechanism has recently been coined polariton-assisted photon recycling (see Fig. \ref{fig:recycling}), and has a weak coupling analogue in luminescent solar concentrators. There, high-energy photons initially absorbed by luminescent molecules can be re-absorbed by other molecules within the concentrator, leading to multiple cycles of absorption and re-emission, ultimately resulting in a red-shifted photoluminescence~\cite{Batchelder1981concentrators,Lindsey2010concentrators,Balaban2014concentrator}. In the strong coupling case, this can occur despite the molecules being far from each other. Indeed, one of the first applications of this phenomenon in the molecular polaritons field can be traced back to seminal works on polariton-assisted remote energy transfer (PARET)~\cite{coles2014polariton,Feist2020ET,du2018theory}. More recently, photon recycling has been proposed to excite molecules at a frequency lower than that of the original pump~\cite{sokolovskii2024photochemical}, and to explain changes in the thermally activated delayed fluorescence kinetics of multi-resonance emitters~\cite{Tomo2024tadf}. Many other interesting polaritonic effects on photoluminescence have recently been reported in which this photon recycling mechanism may play a role~\cite{Borjesson2022interplay,satapathy2021selective,julia2024blue}.



\subsection{\label{sec:vr}Vibrational relaxation}

Some experimental studies report different dark-state-to-polariton relaxation mechanism. The mechanism describes the re-population of the polaritons as incoherent excitons scattering into the polaritons modes via the emission of a vibrational quantum, a process that is mediated by the matter component of the polaritons. This vibrational relaxation process cannot be explained as an optical filtering effect~\cite{chovan2008controlling,coles2011vibrationally,somaschi2011ultrafast}. Seminal works by Litinskaia \textit{et al.}~\cite{litinskaia2004fast} proposed the vibrational relaxation mechanism as an alternative to radiative pumping, potentially becoming a preferential relaxation channel for molecules where non-radiative processes dominate over photoluminescence. Later, experimental works claimed that this channel can occur within timescales as short as hundreds of femtoseconds~\cite{somaschi2011ultrafast}. Over the years, the vibrational relaxation mechanism has remained something of a mystery. On the one hand, theoretical and experimental works have shown this rate to be noncompetitive with radiative pumping, possibly due to the large number of molecules in the system~\cite{del2015quantum,martinez2019triplet,grant2016efficient,Zaumseil2021SWCN}. Thus, these works suggest that vibrational relaxation plays an insignificant role in the dark-state-to-polariton relaxation process in the large-$N$ limit.  On the other hand, a different set of computational and experimental works alludes to a mechanism called vibrationally-assisted scattering (VAS) that does play a significant role when creation of vibrational excitations are involved. Although in some cases this mechanism can be clearly identified as fluorescence \cite{mazza2013microscopic}, other works show that relaxation from dark states to polaritons involves \textit{sharp} features that coincide with the frequency of some Raman-active modes \cite{coles2011vibrationally,somaschi2011ultrafast}. More recently, Tichauer \textit{et al.} conducted computational studies using semiclassical nonadiabatic molecular dynamics simulations to investigate the mechanism behind vibrational relaxation. Their findings revealed a continuous, rather than sharp, population transfer between dark states and polariton states~\cite{Tichauer2021multiscale}, and showed that some Raman-active modes do not contribute to the process~\cite{tichauer2022identifying}. These seemingly contradictory results highlight the need for a unified analytical theory of radiative pumping and vibrational relaxation.
\begin{figure}[!h]
\centering
\includegraphics[width=\linewidth]{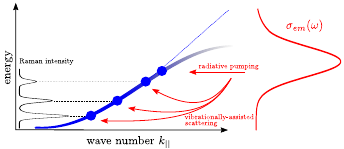}
\captionof{figure}{\textit{Mechanism of vibrationally-assisted scattering.--} Relaxation from dark to polariton states is facilitated by scattering of polaritons whose energy difference coincides with Raman-active modes.}\label{fig:vib_relax}
\end{figure}

Another problem lies in the distinction between these two mechanisms: radiative pumping relies on the photoluminescence spectrum overlapping with the polaritons, while vibrational relaxation depends on the alignment of the frequency difference between polaritons and dark states with the frequencies that correspond to Raman-active modes (see Fig. \ref{fig:vib_relax}). However, low-frequency tails of the photoluminescence spectrum can always overlap with any polariton mode, so radiative pumping cannot be easily ruled out as the mechanism in play~\cite{McGhee2022polcond,groenhof2019tracking}. 

The vibrational relaxation rate can be derived within the same framework used to derive the radiative pumping rate (see Sec.~\ref{sec:rp})~\cite{Perez2025radiative}. The result reveals that the vibrational relaxation rate in the weak vibronic coupling regime originally calculated by Litinskaia \textit{et al.} (proportional to $1/N$)~\cite{litinskaia2004fast} consists only of the rate of radiative pumping from the low-frequency tails of the photoluminescence. Yet, a more careful analysis shows that vibrational relaxation also includes a smaller contribution from a second-order processes in $g$ that we call polariton-assisted Raman scattering, whose rate is proportional to $1/N^2$. This scattering mechanism corresponds to the virtual emission from an incoherent exciton and subsequent spontaneous Raman scattering by a second molecule that results in the creation of a polariton at lower frequency (since the light emitted creates both a Raman phonon and a polariton)~\cite{Perez2025radiative}. Finally, notice that this effect proceeds via the photonic component of the polariton mode~\cite{koner2024nonlinear}.

We argue that polariton-assisted Raman scattering may be responsible for the discrete features in the photoluminescence attributed to VAS, and find no evidence that it can outcompete the radiative pumping rate due to its $1/N^{2}$ dependence. Although it is believed that scattering can dominate over radiative pumping if there is no overlap between the bare emission and the polariton transmission, the frequencies of the Raman-active modes that would allow relaxation via scattering also give rise to shoulders at the low-energy tails of the bare emission spectrum. This implies that radiative pumping will be in competition with polariton-assisted Raman scattering in being the dominant relaxation mechanism that populates the lowest-frequency polariton states. One possible resolution to this conundrum would be that Raman processes are enhanced in multimode cavities due to resonant and off-resonant Raman scattering mediated by the entire lower polariton band~\cite{Perez2025radiative}. Hence, examining the competition between radiative pumping, polariton-assisted photon recycling, and polariton-assisted Raman scattering in realistic mirocavities is crucial. In particular, understanding the extent to which these mechanisms play a role in long-range polariton transport would be valuable.


\subsection{Large number of excitations}

The model presented in Section ~\ref{theory} only considers the first excitation manifold, \textit{i.e.}, $N_{ex}/N\rightarrow 0$, where $N_{ex}$ is the number of excitations per cavity mode. In this limit, it is highly unlikely for a single molecule to be excited more than once. This is based on the assumption that the probe laser is of low intensity, as is the internal (circulating) field of the cavity. If either of these conditions is violated, two interesting situations may occur: (1) higher-order molecular susceptibilities, \textit{i.e.}, $\chi^{(i>1)}(\omega)$, become important and (2) phenomena that were of $\mathcal{O}(1/N)$ in the first excitation manifold can now be magnified. One relevant example of the latter is the radiative pumping rate into an exciton-polariton condensate. The radiative decay is filtered by the polaritons and stimulated by the number of excitations~\cite{mazza2013microscopic,grant2016efficient,cortese2017collective,pannir2022driving}, whose rate now becomes of $\mathcal{O}(N_{exc}/N)$. Notably, in the limit of $N_{exc}\sim N$, photoluminescence can be faster than the dephasing rate, and single-molecule couplings processes can no longer be considered perturbatively. 

\subsection{Few-molecule strong coupling}

In scenarios where strong coupling is achieved with small $N$, higher-order correlations that go as $\mathcal{O}(N^{-1/2})$ can no longer be disregarded (see Section~\ref{sec:impurity_model}), leading to effects which go beyond the linear optics perspective laid out in this review. This not only includes the polariton-assisted Raman scattering and photon recycling mechanisms discussed in Sec.~\ref{sec:beyond_filtering}, but also higher-order processes in $g$. Promising environments for exploring this regime are plasmonic nano- and picocavites, where the small mode volumes allow for strong coupling with a single molecule or with a few molecules~\cite{zengin2015realizing,chikkaraddy2016single,benz2016single,santhosh2016vacuum}. Small mode volumes can also be achieved with surface phonon polaritons to reach vibrational strong coupling within the small-$N$ limit \cite{hirshmann2024spatially}. There are also specialized optical setups that do not require tiny mode volumes to achieve single- and few-molecule strong coupling ~\cite{vahid2021single,koner2023path}.

\subsection{Other polaritonic phenomena beyond optical filtering}

Similar to VAS on femtosecond timescales \cite{somaschi2011ultrafast}, nonlinear spectroscopy experiments of polaritonic systems continue to reveal effects that do not seem to be explainable via optical filtering arguments in the photochemistry and photophysics of systems under strong coupling. For example, Ref. \citenum{xu2023ultrafast} reports observing fast exciton-polariton transport after non-resonantly pumping the system. As the authors state, such an initial excitation primarily populates the exciton states that are uncoupled to the photonic mode.\cite{xu2023ultrafast} Thus, the observed phenomenon requires ultrafast dark-state-to-polariton relaxation. Other examples of experiments under electronic strong coupling that go beyond optical filtering include enhanced energy transfer between J-aggregated cyanine dyes~\cite{zhong2016non} and  between carbon nanotubes~\cite{son2022energy}. For both experiments, the results are attributed to an energy cascade from the upper- to lower-polariton states; Ref.~\citenum{zhong2017energy} reports a time scale of a few picoseconds, while Ref.~\citenum{son2022energy} reports a time scale of a few hundred femtoseconds. A similar phenomenon is observed under vibrational strong coupling. Ref. \citenum{xiang2020intermolecular} reports enhanced intermolecular vibrational energy transfer (VET) between two initially uncoupled infrared modes upon strongly coupling both modes to a microcavity. The timescale of this enhanced energy transfer is 5 ps, and was initially attributed to the middle polariton mediating VET between the two infrared modes (see Ref. \citenum{xiang2020intermolecular} supplementary material), although recently it has been reported that a Raman active mode may play a role in mediating the enhanced VET~\cite{hirschmann2023role}. Since all the results discussed above go beyond the optical filtering framework presented in this review, they must be due to single-molecule $\mathcal{O}(g)$ light-matter coupling processes. However, one would predict that these processes are far too slow to account for the ultrafast timescales that the observed phenomena occur on, as the number of molecules per cavity mode is estimated to be large for these systems. A possible resolution is that the number of molecules per mode has been overestimated for such systems, although it is not clear then why classical optics is so good at explaining their linear optical spectra. Further theoretical understanding of nonlinear spectroscopy of polaritons is needed to explain how effects beyond optical filtering can be prevalent when $N$ is supposedly large. 


\subsection{Experimental evidence of polaritons going beyond optical filtering}
\begin{figure}[!h]
\centering
\includegraphics[width=1\linewidth]{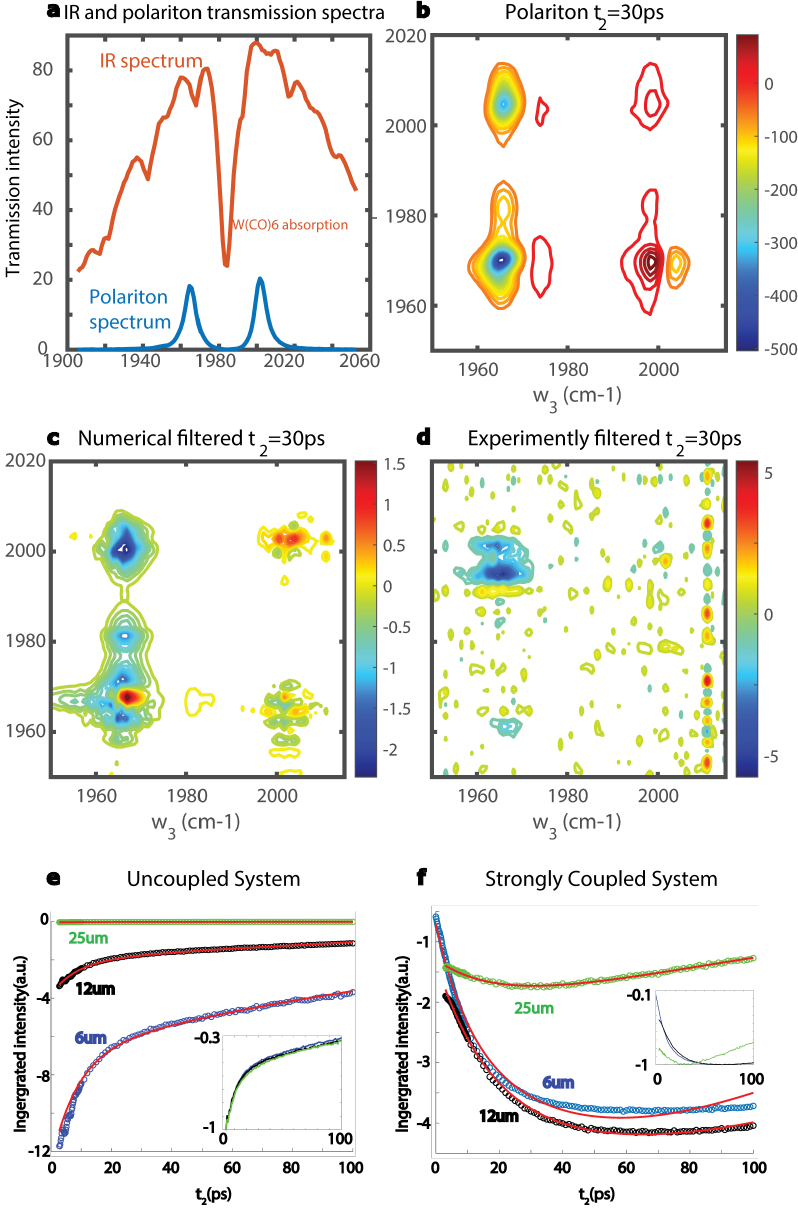}
\captionof{figure}{\textit{Tests of polariton filter effects in 2D IR experiments.--} (a) IR transmission spectra of $\textrm{W(CO)}_6$ in hexane and the same sample inside a polaritonic microcavity. The polariton significantly reduces the amount of IR light transmitted and interacts with the sample as a “filter”. Comparison of 2D IR spectra of (b) the $\textrm{W(CO)}_6$:hexane polariton system, (c) extracavity $\textrm{W(CO)}_6$:hexane filtered numerically by the polariton transmission spectra, and (d) extracavity $\textrm{W(CO)}_6$:hexane filtered experimentally by the polariton transmission spectra. All spectra were scaled to ensure they are compared under the same incidence IR power. $\omega_1$ and $\omega_3$ are the pump and probe frequencies, respectively. The filtered spectra from uncoupled molecules show much smaller intensity. (e) The $\nu = 1 \to 2$ vibrational dynamics of $\textrm{W(CO)}_6$ in hexane outside of the cavity exhibit clearly different signal magnitudes, but when normalized (inset), they demonstrate identical dynamics. (f) The dynamics of $\nu = 1 \to 2$ transitions of the dark reservoir modes at $\omega_{LP}$ show a qualitatively different dependence on cavity thickness, suggesting that the dynamics of strongly coupled systems are not dominated by the free space molecular dynamics (Reproduced from Ref.~\citenum{Simpkins2023comment}, Copyright 2023 B. Simpkins et al.,published by the American Chemical Society. Licensed  under CC-BY 4.0. To view a copy of this license, visit https://creativecommons.org/licenses/by/4.0/.)\label{fig:wei_figure1}}
\end{figure}

\begin{figure}[!ht]
\centering
\includegraphics[width=1\linewidth]{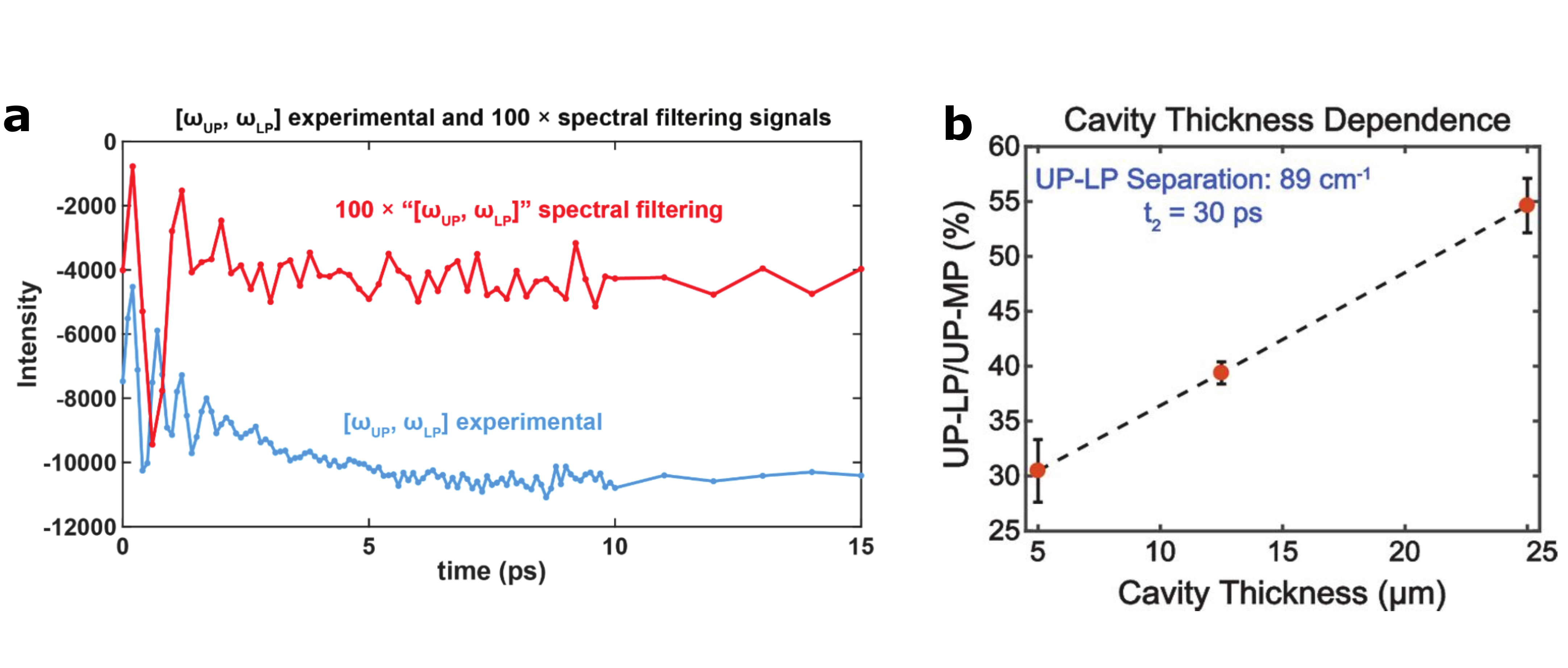}

\captionof{figure}{\textit{Beyond optical filtering effects.--} (a) Experimental peak intensities for strongly coupled $\textrm{Fe(CO)}_5$ inside a cavity (blue) vs 100 times of the intensities of corresponding peaks after spectral filtering is applied to the spectrum for $\textrm{Fe(CO)}_5$ outside the cavity (red) (b) Plot of $I_{UP,MP}/I_{UP,LP}$ as a function of cavity thickness at $\textrm{t}_2 = 30$ ps. Error bars represent the standard deviation of three independent scans (Fig. 14a reproduced from Ref.~\citenum{chen2022cavityenabled}  with permission from AAAS, Copyright 2022. Fig. 14b reproduced from Ref. \citenum{xiang2020intermolecular} with permission from AAAS, Copyright 2020.)\label{fig:wei_figure2}}
\end{figure}

Although some experimental results can be interpreted as a linear filtering effects, notably, some results in ultrafast dynamics do not seem to be simply attributable to this phenomenon. For example, in an effort to compare the two-dimensional infrared (2D IR) spectra of polaritons and the polariton “filtered” 2D IR spectra of molecular species, Simpkins and colleagues juxtaposed the 2D IR spectra of polaritons with spectra that were filtered using two different methods \cite{Simpkins2023comment} (Fig.~\ref{fig:wei_figure1}a-d). First, in a method referred to as ``numerical filtering,'' the authors multiplied the 2D IR spectra of molecular $\textrm{W(CO)}_6$ by the corresponding polariton spectra along both the pump and probe axes (Fig.~\ref{fig:wei_figure1}c). Secondly, in the ``experimental filtering'' method, they used a pulse shaper to create pump spectra mimicking the polariton spectral shape, while the probe was sent directly through a polariton sample as the filter. Both beams were focused onto the same molecular polaritonic sample (Fig.~\ref{fig:wei_figure1}d). Clearly, neither numerical nor experimental filtering can fully reproduce polaritonic 2D IR spectral lineshapes (Fig.~\ref{fig:wei_figure1}b). A possible resolution to this inconsistency is that neither of these two methods considers both pump and probe pulses with the same frequency-dependent \textit{phases} as the intracavity fields (not just their frequency profiles). As discussed in Sec.\ \ref{sec:stcd}, phases must be considered carefully to reproduce transient effects caused by one-photon phase control.


Simpkins and coworkers also compared the ultrafast vibrational dynamics of $\textrm{W(CO)}_6$ both outside cavity and under strong coupling conditions~\cite{Simpkins2023comment}. They observed two significant differences. First, while in the case of the outside of the cavity, the $1 \to 2$ transition intensity simply decayed exponentially in time, reflecting a relaxation to the ground vibrational state (Fig.~\ref{fig:wei_figure1}e), the corresponding transition under strong coupling exhibited a signature rise and decay dynamics (Fig.~\ref{fig:wei_figure1}f), attributed to either the involvement of higher excited vibrational levels~\cite{xiang2019stateselective,xiang2021molecular,li2021cavity,ribeiro2021enhanced} or Raman active modes~\cite{hirschmann2023role}. Markedly, the vibrational dynamics of $\textrm{W(CO)}_6$ with and without strong coupling manifest differently. Secondly, while the dynamics of bare molecular systems are independent of the sample thickness (Fig.~\ref{fig:wei_figure1}e), the strongly coupled system shows a large dependence on the cavity thickness, indicating that the cavity plays an active role in the polariton dynamics (Fig.~\ref{fig:wei_figure1}f).

Chen, Du, Yang and co-workers also compared polariton dynamics against optical filtering controls and showed that the two situations differ dramatically~\cite{chen2022cavityenabled} (Fig.~\ref{fig:wei_figure2}a). In another work, Xiang and co-workers reported polariton enabled intermolecular vibrational energy transfer, and remarkably found that the spectral cross-peak ratio, an indicator of the efficiency of energy transfer,~\cite{xiang2020intermolecular} increased with the cavity thickness (Fig.~\ref{fig:wei_figure2}b) -- again, hinting that the cavity plays an active role in controlling molecular polariton dynamics. 

These outlier results of the linear response theory strongly suggest that other mechanisms beyond optical filtering effects may be takeing place in these dynamic measurements, which are exclusive to cavity strong coupling phenomenona. There have been discussions of higher order nonlinear excitations~\cite{xiang2019stateselective,xiang2021molecular,li2021cavity,ribeiro2021enhanced} or involvement of Raman excited states through coupling to low frequency phonon modes;~\cite{hirschmann2023role} both of these explanations rest on nonlinear interactions which may become available inside of cavities. 


\section{Conclusion and outlook}

This review explores the optical-filtering behavior of molecular polaritons within the collective regime, focusing on the implications of the large-$N$ limit. In this limit, the molecular absorption inside a cavity depends on the cavity quality factor and spectral overlap between the polariton linear transmission and the bare molecular absorption. We re-evaluate several predicted polaritonic phenomena in this context and suggest that as $N\to \infty$, polaritons function in large part as optical filters, selectively allowing certain frequencies to enter the cavity that are then absorbed by the molecules. Through the theoretical and experimental examples presented, we demonstrate how this perspective can straightforwardly account for various polaritonic phenomena proposed in the literature and that similar outcomes should be replicable outside the cavity using appropriately tailored light sources. With these findings in mind, we urge caution in attributing phenomena solely to ``polaritonic'' or to quantum effects without first thoroughly accounting for classical optical filtering effects.

The linear optical treatment of polaritons is \textit{exact} in the asymptotic $N\to\infty$ limit. However, this picture breaks down as $\mathcal{O}(g)$ quantum processes, or ``$1/N$ effects'', become significant. Examples of these $1/N$ effects are dark-state-to-polariton relaxation, fluorescence, and spontaneous Raman scattering. Similarly, in systems under strong coupling with a small number of molecules or with many quanta of excitation, quantum optics effects cannot be ignored, further limiting the scope of the linear optical perspective. Although quantum mechanical in nature, some $1/N$ effects like fluorescence can be described using classical linear optics. However, in the strong coupling regime, they can only partially be described as solely optical filtering due to phenomena like polariton-assisted photon recycling. Finally, there is experimental evidence for ultrafast behavior of polaritonic systems in the collective regime that theory cannot yet reconcile without invoking single-molecule processes that are artificially large. Thus, for both theorists and experimentalists aiming to uncover quantum or classical polaritonic phenomena that transcend the realm of optical filtering, the focus should shift towards these regions of interest. This approach will not only expand our understanding of polaritonic effects beyond the linear regime but also pave the way for novel applications and insights into the complex interplay between light and matter under strong coupling.

\section*{Conflicts of interest}
There are no conflicts to declare.

\section*{Data availability}
The authors confirm that the parameters and equations used in the calculations carried out to support the conclusions of this study are available within the review article and cited works.

\section*{Acknowledgements}

K.S., A.K., J.B.P.-S, W.X., N.C.G, and J.Y.-Z. acknowledge support by the Air Force Office of Scientific Research (AFOSR) through the Multi-University Research Initiative (MURI) program no. FA9550-22-1-0317. M.L.W. acknowledges support from the National Science Foundation CAREER program under award CHE-2238865 and the US Department of Energy, Office of Science, Basic Energy Sciences, CPIMS Program under Early Career Research Program award DE-SC0022948. The authors thank Abraham Nitzan for pointing out the analogies between collective vs single-molecule couplings with Rayleigh and Raman scattering amplitudes, respectively.


\balance


\providecommand*{\mcitethebibliography}{\thebibliography}
\csname @ifundefined\endcsname{endmcitethebibliography}
{\let\endmcitethebibliography\endthebibliography}{}

\end{document}